\documentclass[a4paper,11pt]{article}
\usepackage{jheppub} 
\makeatletter
\def\@fpheader{}
\makeatother
\usepackage{amsmath,mathrsfs,braket,hyperref,bm,amssymb,graphicx}
\usepackage {slashed}
\DeclareMathOperator{\Tr}{Tr}

\usepackage{tikz}
\usepackage{tikz-feynman}
\usepackage{subcaption}  

\numberwithin{equation}{section}
\allowdisplaybreaks[4]

\newcommand{\gammaE}{{\gamma_\rmii{E}}}

\newcommand{\be}{\begin{equation}}
	\newcommand{\ee}{\end{equation}}
\newcommand{\lp}{\left(}
\newcommand{\rp}{\right)}
\newcommand\diff{\,\mathrm{d}}

\newcommand{\nn}{\nonumber}

\newcommand{\rmii}[1]{{\mbox{\tiny\rm{#1}}}}

\renewcommand{\Re}{\mathrm{Re}}
\renewcommand{\Im}{\mathrm{Im}}

\title{\boldmath Nonequilibrium framework for first-order phase transitions }

\author[a]{Jie Liu,}
\emailAdd{jieliu@stu.cqu.edu.cn}

\author[a]{Ligong Bian\footnote{Corresponding Author},}
\emailAdd{lgbycl@cqu.edu.cn}
\affiliation[a]{Department of Physics and Chongqing Key Laboratory for Strongly Coupled Physics, Chongqing University, Chongqing 401331, P. R. China}

\begin{document}
	
\abstract{The first-order phase transition is central to electroweak baryogenesis and gravitational wave generation, yet thermal equilibrium effective potential analyses cannot describe the real-time non-equilibrium dynamics of the phase transition process. We use the Schwinger-Keldysh closed-time-path formalism of real-time finite-temperature field theory to study these effects. For the $\phi^6$ model, we derive the effective action and the corresponding Langevin equation. We further extend the formalism to the two scalar fields case, obtaining the coupled Langevin equations. These results provide a theoretical framework for the bubble nucleation simulation of the non-equilibrium phase transition process, laying the foundation for reliable predictions of baryogenesis and gravitational wave signals.}

\maketitle
\flushbottom	
\section{Introduction}
\label{sec:intro}
The electroweak phase transition (PT) provides an important window into the thermal history of the early Universe and possible new physics around the electroweak scale. A strong first-order electroweak PT can provide the departure from thermal equilibrium required by the Sakharov conditions for electroweak baryogenesis~\cite{Kuzmin:1985mm,Blinov:2015sna,Sakharov:1967dj}, and may also generate a stochastic gravitational wave background potentially accessible to future space-based interferometers~\cite{Kosowsky:1992rz,Kamionkowski:1993fg,Athron:2023xlk}. For the experimentally measured Higgs mass at LHc~\cite{ATLAS:2012yve,CMS:2012qbp}, the electroweak PT in the Standard Model is a smooth crossover~\cite{Kajantie:1996mn}. Realizing a first-order electroweak PT therefore generally requires interactions beyond the Standard Model, among which extensions of the scalar sector provide a simple and widely studied possibility~\cite{Vaskonen:2016yiu,Profumo:2014opa,Kozaczuk:2015owa,Chung:2012vg}.

Traditionally, the electroweak PT is studied based on the effective potential within the imaginary-time formalism of thermal field theory~\cite{Hirvonen:2021zej,Croon:2020cgk,Laine:2016hma,Kajantie:1995dw,Niemi:2021qvp}. Combined with the finite-temperature Euclidean bounce formalism, this approach can be used to calculate the critical temperature, phase transition strength, and nucleation temperature, and to study the associated gauge dependence and gravitational wave. However, this framework mainly describes the equilibrium free energy and the Euclidean tunneling process, which does not directly capture the real-time nonequilibrium evolution of the field during the PT. In real time, the background fields continuously interact with the thermal environment, leading to dissipation and stochastic thermal fluctuations~\cite{Calzetta_Hu_2023,Kapusta_Gale_2023}. Therefore, the conventional effective-potential approach alone does not describe the nonequilibrium dynamics of the PT.

A real-time nonequilibrium formulation is required to describe these effects. The Schwinger--Keldysh closed-time-path (CTP) formalism provides a natural framework for this purpose~\cite{Schwinger:1960qe,Keldysh:1964ud,Chou:1984es,Landsman:1986uw,Lundberg:2020mwu}. The CTP effective action yields real and causal equations of motion for the background fields. After the thermal fluctuations are integrated out, causal nonlocal kernels in the real part of the effective action encode the dissipative response of the background fields to the thermal environment, whereas the imaginary part defines the corresponding noise kernels. By introducing auxiliary stochastic fields, the latter can be represented in terms of stochastic fluctuations. Dissipation and noise can thus be derived consistently from the same microscopic finite-temperature quantum field theory and, in appropriate limits, are related by the fluctuation--dissipation relation, without introducing phenomenological friction or stochastic forces by hand~\cite{Gleiser:1993ea}. 

In this work, We first consider a single scalar field $\phi^6$ model, which can generate a barrier for a first-order PT, may also arise in the low-energy effective theory after heavier new-physics degrees of freedom are integrated out~\cite{Grojean:2004xa,Zhu:2025pht,Liu:2025ipj,Liu:2026ask}. This model offers a convenient setting to study nonequilibrium dynamics and to show explicitly how dissipative terms and stochastic fluctuations arise consistently from the real-time CTP effective action, leading to an effective Langevin equation. We further extend the analysis to two scalar model to mimic the real singlet scalar extended Standard Model~\cite{Niemi:2024axp,Kakizaki:2015wua,Alves:2018jsw,Hashino:2016xoj,Chiang:2018gsn,Profumo:2007wc,Espinosa:2011ax,Kurup:2017dzf}, where the Higgs field and the additional real singlet form a coupled two-field system. This model has a richer finite temperature vacuum structure and can realize one-step or two-step PTs~\cite{Niemi:2020hto,Bian:2018bxr}. The additional singlet scalar also be related to dark matter phenomenology~\cite{Cline:2013gha,Cline:2012hg}. Compared with the single-field model, the two-field system further introduces coupled dissipation between different order parameters and correlations between different stochastic sources, leading to a richer nonequilibrium dynamics. Under suitable locality and Markovian approximations~\cite{Morikawa:1986rp}, the resulting nonlocal dissipative terms can be reduced to effective local dissipation coefficients.

The paper is organized as follows. In Sec.~\ref{sec:effective action} we derive the effective action for a nonuniform time-dependent background field configuration $\varphi(\bm{x},t)$, up to two loops and order $\lambda^2$ and $c_6$. In Sec.~\ref{sec:effective equation of motion} we obtain the effective equation of motion for $\varphi(\bm{x},t)$ and discuss the approximations involved in order that it obeys a Langevin-Like equation. In Sec.~\ref{sec:two field model} we extend the above CTP calculation to the two-field case, obtaining the complete coupled Langevin equations that describe the nonequilibrium evolution of the two-field system. The summary is presented in Sec.~\ref{sec:conclusion}. Finally, the Appendixes are included in order to present some technical results used in the paper.

\section{The effective action}\label{sec:effective action}

\begin{figure}[!htp]
    \centering
    \includegraphics[width=0.8\linewidth]{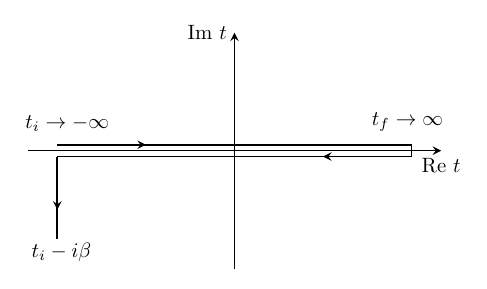}
    \caption{The closed time-path contour.}
    \label{fig:Closed time-path contour}
\end{figure}

Considering the scalar field model with the Lagrangian density
\begin{equation}
    \mathcal{L}[\phi]=\frac{1}{2}(\partial_\mu \phi)^2-V(\phi)\;,
\end{equation}
the scalar potential is
\begin{equation}
    V(\phi)=\frac{1}{2} m^2 \phi^2+\frac{\lambda}{4!} \phi^4+\frac{c_6}{6!}\phi^6\;,
\end{equation}
and the generating functional $Z[J]$, in terms of an external source $J$, is given by
\begin{equation}
    Z[J]=\int_c \mathcal{D}\phi \exp{iS[\phi,J]}\;,
\end{equation}
where the suffix c represents the closed time contour of integration  depicted in Figure~\ref{fig:Closed time-path contour} and the classical action is given by 
\begin{equation}
    S[\phi,J]=\int_c \diff^4 x \{\mathcal{L}[\phi]+J(x)\phi(x) \}.
\end{equation}

The effective action $\Gamma[\varphi]$ is obtained from the Legendre transformation of the connected generating functional $W[J]$,
\begin{equation}
    \Gamma[\varphi]=W[J]-\int_c \diff^4 x J(x)\varphi(x)\;,
\end{equation}
where
\begin{equation}
    \varphi(x)=\frac{\delta W[J]}{\delta J(x)}\;,
\end{equation}
and the function $W$ is
\begin{equation}
    W[J]=-i\ln{Z[J]}\;.
\end{equation}
The detailed derivation of the perturbative loop expansion for $\Gamma[\varphi]$ is provided in Appendix~\ref{appendix:Gamma}.
We now consider the contributions from this expansion up to two-loop order, to order $\lambda^2$ and $c_6$. Graphically we have
\begin{align}
\Gamma[\varphi] &= S[\varphi] \nn\\[5pt]
& +\;
\begin{tikzpicture}[baseline=0pt,thick] 			
	\begin{scope}[xshift=0.8cm]
		\draw (0,0) circle (0.4);
		\draw (-0.4,0) -- (-0.8,0.4);
		\draw (-0.4,0) -- (-0.8,-0.4);
	\end{scope}
\node at (1.5,0) {$+$};			
	\begin{scope}[xshift=2.6cm]
	    \draw (0,0) circle (0.4);
	    \draw (-0.4,0) -- (-0.8,0.4);
	    \draw (-0.4,0) -- (-0.8,-0.4);
	    \draw (0.4,0) -- (0.8,0.4);
	    \draw (0.4,0) -- (0.8,-0.4);
	\end{scope}			
\node at (3.6,0) {$+$};		
	\begin{scope}[xshift=5.1cm]
		\draw (-0.4,0) circle (0.4);
		\draw (0.4,0) circle (0.4);
		\draw (-0.8,0) -- (-1.2,0.4);
		\draw (-0.8,0) -- (-1.2,-0.4);
	\end{scope}
\node at (6.2,0) {$+$};
	\begin{scope}[xshift=7.2cm]
		\draw (-0.8,0) -- (0.8,0);
		\draw (0,0) circle (0.4);
	\end{scope}
\end{tikzpicture} \nn\\[4pt]  
& + \;
\begin{tikzpicture}[baseline=0pt,thick]
	\begin{scope}[xshift=0.8cm]
		\draw (-0.8,0) -- (0.8,0);
		\draw (0,0.4) circle (0.4);
		\draw (0,0) -- (-0.4,-0.4);
		\draw (0,0) -- (0.4,-0.4);
	\end{scope}
\node at (1.8,0) {$+$};	
	\begin{scope}[xshift=2.8cm]
		\draw (-0.8,0) -- (0.8,0);
		\draw (0,0.4) circle (0.4);
		\draw (0,-0.4) circle (0.4);
	\end{scope}
\node at (4.7,0) {$+\;\mathcal{O}(\lambda^3,c_6^2)$\;,};
\end{tikzpicture}\label{eq:action picture}
\end{align}
where, in the graphic representation, $\varphi$ is in the external legs and the internal propagators are the real-time finite-temperature propagators $G_{\phi}^{n,l}(n,l=+,-)$.

Within the CTP formalism, based on the thermal density matrix $\rho$~\cite{Landsman:1986uw}, the four components of $G_{\phi}^{n,l}(n,l=+,-)$ can be expressed via the expectation values of the field operators as
\begin{align}
	G_{\phi}^{++}(x-x')&=\braket{T_+\phi(x)\phi(x')}=\Tr[\rho T_+\phi(x)\phi(x')]\,,\nn\\
	G_{\phi}^{--}(x-x')&=\braket{T_-\phi(x)\phi(x')}=\Tr[\rho T_-\phi(x)\phi(x')]\,,\nn\\
	G_{\phi}^{+-}(x-x')&=\braket{\phi(x)\phi(x')}=\Tr[\rho\phi(x)\phi(x')]\,,\nn\\
	G_{\phi}^{-+}(x-x')&=\braket{\phi(x')\phi(x)}=\Tr[\rho\phi(x')\phi(x)]\,.
\end{align}
Here, $T_+$ and $T_-$ denote the time-ordering operators along the forward and backward branches of the time contour respectively and $\rho=2\pi\mathrm{sgn(k_0)\delta(k_0^2-\omega^2)}$ for the free scalar particle. Fourier transforming these matrix elements into momentum space yields the explicit integral representation of the propagator matrix:
\begin{align}
	G_{\phi}(x-x')=\int \frac{\diff^3 k}{(2\pi)^3}e^{-i \bm{k}(x-x')} \begin{pmatrix}
	G_{\phi}^{++}(\bm{k},t-t') & G_{\phi}^{+-}(\bm{k},t-t')\\
	G_{\phi}^{-+}(\bm{k},t-t') & G_{\phi}^{--}(\bm{k},t-t')
	\end{pmatrix}\;,
\end{align} 
where
\begin{align}
    G_{\phi}^{++}(\bm{k},t-t')=&\theta(t-t')G_{\phi}^{>}(\bm{k},t-t')+\theta(t'-t)G_{\phi}^{<}(\bm{k},t-t')\nn\;,\\
	G_{\phi}^{--}(\bm{k},t-t')=&\theta(t'-t)G_{\phi}^{>}(\bm{k},t-t')+\theta(t-t')G_{\phi}^{<}(\bm{k},t-t')\nn\;,\\
	G_{\phi}^{+-}(\bm{k},t-t')=&G_{\phi}^{>}(\bm{k},t-t')\nn\;,\\
	G_{\phi}^{-+}(\bm{k},t-t')=&G_{\phi}^{<}(\bm{k},t-t')\;.
\end{align}
And, for a free scalar field at finite temperature, the explicit Fourier transforms of these Wightman functions are given by
\begin{align}\label{eq:Wightman functions}
    G_{\phi}^{>}(\bm{k},t-t')=&\frac{1}{2 \omega(\bm{k})}\Big\{\big[ 1+2 n(\omega) \big]\cos[\omega(t-t')]-i\sin[\omega(t-t')]\Big\}\nn\,,\\
	G_{\phi}^{<}(\bm{k},t-t')=&G_{\phi}^{>}(\bm{k},t'-t)\;.
\end{align}
Where, $n(\omega)=\lp e^{\beta \omega}-1 \rp^{-1}$ is the Bose distribution and $\omega \equiv \omega(\bm{k})=\sqrt{\bm{k}^2+m^2}$ is the free particle energy.

In terms of the field variables $\varphi_+$ and $\varphi_-$, the first term in Eq.~\eqref{eq:action picture} is given by
\begin{equation}
    S[\varphi]=\int \diff^4x\left\{\mathcal{L}[\varphi_+]-\mathcal{L}[\varphi_-] \right\}\;.
\end{equation}
The first graph's contribution reads
\begin{align}
\begin{picture}(90,0)
\thicklines
	\put(0,5){\line(-1,-1){10}}
	\put(10,5){\circle{20}}
	\put(0,5){\line(-1,1){10}}
\end{picture}
\hspace{-2.4cm}&=-\frac{\lambda}{4}\int \diff^4 x\int \frac{\diff^3 q}{(2 \pi)^3}\bigg\{ \varphi_+^2(x) G_{\phi}^{++}(\bm{q},0)-\varphi_-^2(x) G_{\phi}^{--}(\bm{q},0)\bigg\}\nn\\
&=-\frac{\lambda}{4}\int \diff^4 x\Big[\varphi_+^2(x)-\varphi_-^2(x)\Big]\int\frac{d^3 q}{(2 \pi)^3}\frac{1}{2 \omega(\bm{q})} \left[1 + 2 n(\omega)\right]\;,
\end{align}
The second graph's contribution is given by
\begin{align}
\begin{picture}(200,0)
	\thicklines
	\put(0,4){\line(-1,-1){10}}
	\put(10,4){\circle{20}}
	\put(0,4){\line(-1,1){10}}
	\put(20,4){\line(1,1){10}}
	\put(20,4){\line(1,-1){10}}
\end{picture}
\hspace{-6cm}
=&i \frac{\lambda^2}{16} \int d^4 x d^4 x' \int \frac{d^3 k}{(2 \pi)^3} e^{i \bm{k} . (\bm{x} - \bm{x'})} \int \frac{d^3 q}{(2 \pi)^3} \left[ \varphi_{+}^2 (x) G_{\phi}^{++}(\bm{q} , t-t')G_{\phi}^{++}(\bm{q}- \bm{k} , t-t') \varphi_{+}^2 (x') \right. \nonumber \\
&-\left. \varphi_{+}^2 (x) G_{\phi}^{+-} (\bm{q} , t-t') G_{\phi}^{+-}(\bm{q} -\bm{k} ,t-t') \varphi_{-}^2 (x')- \varphi_{-}^2 (x) G_{\phi}^{-+}(\bm{q} , t-t')G_{\phi}^{-+}(\bm{q} -\bm{k} , t-t')\varphi_{+}^2 (x') \right.\nonumber \\
&\left. + \varphi_{-}^2 (x) G_{\phi}^{--}(\bm{q} , t-t')G_{\phi}^{--}(\bm{q} -\bm{k} ,t-t') \varphi_{-}^2 (x') \right] \: . 
\end{align}
Equivalently, the third graph's contribution is given by
\begin{align}
\begin{picture}(200,0)
\thicklines
\put(0,5){\line(-1,-1){10}}
\put(10,5){\circle{20}}
\put(30,5){\circle{20}}
\put(0,5){\line(-1,1){10}}
\end{picture}
\hspace{-5.5cm}=&i\frac{\lambda^2}{8}\int \frac{\diff^3 k}{(2 \pi)^3}\frac{1+2 n(\omega)}{2 \omega(\bm{k})}\int \diff^4 x\int \diff t' \frac{\diff^3 q}{(2 \pi)^3}\left\{\varphi_+^2(x)\left[G_{\phi}^{++}(\bm{q},t-t')\right]^2\right.\nn\\
&\left.-\varphi_+^2(x)\left[G_{\phi}^{+-}(\bm{q},t-t')\right]^2-\varphi_-^2(x)\left[G_{\phi}^{-+}(\bm{q},t-t')\right]^2+\varphi_-^2(x)\left[G_{\phi}^{--}(\bm{q},t-t')\right]^2\right\}\;.
\end{align}
Similarly, we can get the expression of the fourth graph in Eq.~\eqref{eq:action picture} as:
\begin{align}
\begin{picture}(200,0)
\thicklines
\put(0,5){\line(1,0){40}}
\put(20,5){\circle{20}}
\end{picture}
\hspace{-5.5cm}=&
i\frac{\lambda^2}{12}\int \diff^4 x \diff^4 x' \int \frac{\diff^3 k}{(2 \pi)^3} e^{i \bm{k}.(\bm{x}-\bm{x'})}\int \frac{\diff^3 q_1}{(2 \pi)^3} \frac{\diff^3 q_2}{(2 \pi)^3} \frac{\diff^3 q_3}{(2 \pi)^3} \delta(\bm{k}-\bm{q}_1-\bm{q}_2-\bm{q}_3)\nn\\
&\times \bigg\{\varphi_+(x)G_{\phi}^{++}(\bm{q}_1,t-t')G_{\phi}^{++}(\bm{q}_2,t-t')G_{\phi}^{++}(\bm{q}_3,t-t')\varphi_+(x')\nn\\
&-\varphi_+(x)G_{\phi}^{+-}(\bm{q}_1,t-t')G_{\phi}^{+-}(\bm{q}_2,t-t')G_{\phi}^{+-}(\bm{q}_3,t-t')\varphi_-(x')\nn\\
&-\varphi_-(x)G_{\phi}^{-+}(\bm{q}_1,t-t')G_{\phi}^{-+}(\bm{q}_2,t-t')G_{\phi}^{-+}(\bm{q}_3,t-t')\varphi_+(x')\nn\\
&+\varphi_-(x)G_{\phi}^{--}(\bm{q}_1,t-t')G_{\phi}^{--}(\bm{q}_2,t-t')G_{\phi}^{--}(\bm{q}_3,t-t')\varphi_-(x')\bigg\}\;.
\end{align}
Following the same procedure, the fifth graph in Eq.~\eqref{eq:action picture} is given by
\begin{align}
\begin{picture}(110,10)
\put(-20,0){\line(1,0){40}}
\put(0,0){\line(-1,-1){10}}
\put(0,0){\line(1,-1){10}}
\put(0,10){\circle{20}}
\end{picture}
\hspace{-3cm}&=-\frac{c_6}{48}\int \diff^4 x\int \frac{\diff^3 q}{(2 \pi)^3}\bigg\{ \varphi_+^4(x) G_{\phi}^{++}(\bm{q},0)-\varphi_-^4(x) G_{\phi}^{--}(\bm{q},0)\bigg\}\nn\\
&=-\frac{c_6}{48}\int \diff^4 x \Big[ \varphi_+^4(x)-\varphi_-^4(x)\Big]\int \frac{d^3 q}{(2 \pi)^3}\frac{1}{2 \omega(\bm{q})} \left[ 1 + 2 n(\omega)\right]\;.
\end{align}
Analogously, the last graph in Eq.~\eqref{eq:action picture} is given by
\begin{align}
\begin{picture}(110,20)
\put(0,5){\line(1,0){40}}
\put(20,15){\circle{20}}
\put(20,-5){\circle{20}}
\end{picture}
\hspace{-2.4cm}&=-\frac{c_6}{16}\int \diff^4 x\int \frac{\diff^3 k}{(2 \pi)^3} \frac{\diff^3 q}{(2 \pi)^3}\bigg\{ \varphi_+^2(x) G_{\phi}^{++}(\bm{k},0)G_{\phi}^{++}(\bm{q},0)-\varphi_-^2(x) G_{\phi}^{--}(\bm{k},0)G_{\phi}^{--}(\bm{q},0)\bigg\}\nn\\
&=-\frac{c_6}{16}\int \diff^4 x \Big[  \varphi_+^2(x)-\varphi_-^2(x)\Big]\bigg[ \int \frac{d^3 q}{(2 \pi)^3}\frac{1+2 n(\omega)}{2 \omega(\bm{q})}\bigg]^2\;.
\end{align}
Before proceeding, it is advantageous to rephrase the field variables $\varphi_+$ and $\varphi_-$ in terms of the new variables $\varphi_c$ and $\varphi_\Delta$, defined by
\begin{equation}
\varphi_+=\varphi_c+\frac{1}{2}\varphi_\Delta\;,\qquad \varphi_-=\varphi_c-\frac{1}{2}\varphi_\Delta\;.
\end{equation}
This transformation allows us to clearly distinguish between the physical field and the fluctuations in the effective action. Specifically, $\varphi_c$ denotes the physical field, which ``feels" the fluctuations of the system, while $\varphi_\Delta$ is associated with the response field. At the end of the calculation, the physical identification imposes $\varphi_\Delta\to0(\varphi_+=\varphi_-)$. In terms of these new variables, the effective action can be fully expressed using the physical propagator $G^{++}_\phi$, thereby simplifying the evaluation and physical interpretation of the Feynman diagrams.

Summing these contributions, we arrive at our final expression:
\begin{equation}
\begin{aligned}\label{eq:effective action_varphicdelta}
\Gamma[\varphi_\Delta,\varphi_c]=&\int \diff^4 x \Bigg\{\varphi_\Delta(x)\bigg[-\Box-m^2-\frac{\lambda}{2}\int\frac{\diff^3 q}{(2\pi)^3}\frac{1+2 n(\omega)}{2 \omega(\bm{q})}-\frac{c_6}{8}\bigg[ \int \frac{d^3 q}{(2 \pi)^3}\frac{1+2 n(\omega)}{2 \omega(\bm q)}\bigg]^2\\
&-\frac{\lambda^2}{2}\int \diff t'\int\frac{\diff^3 q}{(2\pi)^3}\Im\Big[G_\phi^{++}(\bm{q},t-t')\Big]^2\theta(t-t') \int \frac{\diff^3 k}{(2 \pi)^3}\frac{1+2 n(\omega)}{2 \omega(\bm{k})}\bigg]\varphi_c(x)\\
&-\bigg[ \frac{\lambda}{4!}+ \frac{c_6}{48}\int \frac{\diff^3 q}{(2 \pi)^3}\frac{1+2 n(\omega)}{2 \omega(\bm{q})}\bigg]\Big[4 \varphi_\Delta(x)\varphi_c^3(x)+\varphi_\Delta^3(x)\varphi_c(x)\Big]\\
&-\frac{c_6}{6!}\Big[6\varphi_\Delta(x) \varphi_c^5(x)+5\varphi_\Delta^3(x) \varphi_c^3(x)+\frac{3}{8}\varphi_\Delta^5(x) \varphi_c(x)\Big]\Bigg\}\\
&+\int\diff^4x\int\diff^4x'\int\frac{\diff^3 k}{(2\pi)^3}e^{i \bm{k}\cdot(\bm{x}-\bm{x}')}\Bigg\{-\frac{\lambda^2}{2}\bigg[ \varphi_c^2(x)\varphi_c(x')\varphi_\Delta(x')\\
&+\frac{1}{4}\varphi_\Delta^2(x)\varphi_c(x')\varphi_\Delta(x')\bigg] \int\frac{\diff^3 q}{(2\pi)^3} \Im\Big[ G_{\phi}^{++}(\bm{q},t-t') G_{\phi}^{++}(\bm{q}-\bm{k},t-t')\Big]\theta(t-t')\\
&-\frac{\lambda^2}{3}\varphi_c(x)\varphi_\Delta(x')\Im \Big[ \prod_{j=1}^{3} \int \frac{\diff^3 q_j}{(2\pi)^3}G_{\phi}^{++}(\bm{q}_j,t-t')\Big]\theta(t-t') \delta(\bm{k}-\bm{q}_1-\bm{q}_2-\bm{q}_3) \\
&+i\frac{\lambda^2}{4} \varphi_c(x)\varphi_\Delta(x)\varphi_c(x')\varphi_\Delta(x')\int \frac{\diff^3 p}{(2\pi)^3} \Re\Big[G_{\phi}^{++}(\bm{q},t-t') G_{\phi}^{++}(\bm{q}-\bm{k},t-t')\Big]\\
&+i\frac{\lambda^2}{12}\varphi_\Delta(x) \varphi_\Delta(x')\Re \Big[\prod_{j=1}^{3} \int \frac{\diff^3 q_j}{(2\pi)^3}G_{\phi}^{++}(\bm{q}_j,t-t') \Big] \delta(\bm{k}-\bm{q}_1-\bm{q}_2-\bm{q}_3)\Bigg\}\;.
\end{aligned}
\end{equation}
The last two terms of Eq.~(\ref{eq:effective action_varphicdelta}) give the imaginary contributions to the effective action. We attribute these imaginary terms to the functional integration over the Gaussian fluctuation fields $\xi_1$ and $\xi_2$~\cite{Morikawa:1986rp}. To handle these terms, we need to employ the Hubbard–Stratonovich identity, which for a generic item $K(x,x')$ takes the form
\begin{equation}
    \exp \left[-\frac{1}{2}\int f K f\right]=\int \mathcal{D}\xi \exp \left[-\frac{1}{2}\xi K^{-1}\xi+i \xi f\right]\,.
\end{equation}
Applying that identity to each imaginary term introduces auxiliary fields that linearize the corresponding interactions, we have
\begin{equation}
    \begin{aligned}\label{eq:Gaussian fluctuation}
        \int \mathcal{D}\xi_1 P[\xi_1]\int&\mathcal{D}\xi_2P[\xi_2]\exp\left\{i\int\diff^4[\xi_1(x)\varphi_\Delta(x)\varphi_c(x)+\xi_2(x)\varphi_\Delta(x)] \right\}\\
        =&\exp\bigg\{i\int\diff^4x\diff^4x'\bigg[i\frac{\lambda^2}{4}\varphi_\Delta(x)\varphi_c(x)\Re\Big[G_{\phi}^{++}\Big]^2_{x,x'}\varphi_\Delta(x')\varphi_c(x') \\
        &+i \frac{\lambda^2}{12}\varphi_\Delta(x)\Re\Big[G_{\phi}^{++}\Big]^3_{x,x'}\varphi_\Delta(x')\bigg]\bigg\}\;,
    \end{aligned}
\end{equation}
where $P[\xi_1]$ and $P[\xi_2]$, and the Gaussian probability distributions for $\xi_1$ and $\xi_2$, are given by
\begin{align}
    P[\xi_1]=N_1 \exp \bigg\{-\frac{1}{2}\int\diff^4x\int\diff^4x'\xi_1(x)\bigg(\frac{\lambda^2}{2}\Re\Big[G_{\phi}^{++}\Big]^2_{x,x'}\bigg)^{-1}\xi_1(x')\bigg\}\;,\label{eq:probability distribution1}\\
    P[\xi_2]=N_2 \exp \bigg\{-\frac{1}{2}\int\diff^4x\int\diff^4x'\xi_2(x)\bigg(\frac{\lambda^2}{6}\Re\Big[G_{\phi}^{++}\Big]^3_{x,x'}\bigg)^{-1}\xi_2(x')\bigg\}\;,\label{eq:probability distribution2}
\end{align}
where $N_1$ and $N_2$ are normalization factors, and in Eqs.~(\ref{eq:Gaussian fluctuation})-(\ref{eq:probability distribution2}) we introduce the compact notation
\begin{equation}\label{eq:compact notation1}
    \Big[G_{\phi}^{++}\Big]^2_{x,x'}=\int \frac{\diff^3 k}{(2 \pi)^3} e^{i\bm{k}(\bm{x}-\bm{x'})}\int \frac{\diff^3 q}{(2\pi)^3}G_{\phi}^{++}(\bm{q},t-t') G_{\phi}^{++}(\bm{q}-\bm{k},t-t')\;,
\end{equation}
and 
\begin{equation}\label{eq:compact notation2}
    \Big[G_{\phi}^{++}\Big]^3_{x,x'}=\int \frac{\diff^3 k}{(2 \pi)^3} e^{i\bm{k}(\bm{x}-\bm{x'})}\prod_{j=1}^{3} \int \frac{\diff^3 q_j}{(2\pi)^3}G_{\phi}^{++}(\bm{q}_j,t-t') \delta(\bm{k}-\bm{q_1}-\bm{q_2}-\bm{q_3})\;.
\end{equation}
Thus, using Eq.~(\ref{eq:Gaussian fluctuation}), the Eq.~(\ref{eq:effective action_varphicdelta}) recasts the form of
\begin{equation}
    \exp\left\{i\Gamma[\varphi_\Delta,\varphi_c]\right\}=\int \mathcal{D}\xi_1 P[\xi_1]\int \mathcal{D}\xi_2P[\xi_2]\exp\{iS_{\mathrm{eff}}[\varphi_\Delta,\varphi_c,\xi_1,\xi_2]\}\;,
\end{equation}
where
\begin{equation}\label{eq:Seff_varphicdelta}
    S_\mathrm{eff}[\varphi_\Delta,\varphi_c,\xi_1,\xi_2]=\Re\Gamma[\varphi_\Delta,\varphi_c]+\int \diff^4 x[\xi_1(x)\varphi_\Delta(x)\varphi_c(x)+\xi_2(x)\varphi_\Delta(x)]\;,
\end{equation}
and $\Re\Gamma[\varphi_\Delta,\varphi_c]$ is the real part of Eq.~(\ref{eq:effective action_varphicdelta}). In Eq.~(\ref{eq:Seff_varphicdelta}), the fields $\xi_1$ and $\xi_2$, with probability distributions given by Eqs.~(\ref{eq:probability distribution1})-(\ref{eq:probability distribution2}), respectively, serve as fluctuation sources for the scalar field configuration $\varphi$. $\xi_1$ couples with both the physical field $\varphi_c$ and with the response field $\varphi_\Delta$, leading to a coupled (multiplicative) noise term  $(\varphi_c \xi_1)$ in the equation of motion for $\varphi_c$, while $\xi_2$ gives  origin to an additive noise term. In the next section, we examine the relevance of each of these noise terms in the equation of motion for the physical field $\varphi_c$, and evaluate the dissipation coefficients associated with them.

\section{The effective equation of motion}
\label{sec:effective equation of motion}
The equation of motion for $\varphi_c$ can be obtained through 
\begin{equation}
    \left.\frac{\delta S[\varphi_\Delta, \varphi_c, \xi_1,\xi_2]}{\delta \varphi_\Delta}\right|_{\varphi_\Delta=0}=0\,.
\end{equation}
According to Eqs.~(\ref{eq:Seff_varphicdelta}) and~(\ref{eq:effective action_varphicdelta}), we can get
\begin{equation}
    \begin{aligned}\label{eq:equation of motion}
\Bigg[\Box&+m^2+\frac{\lambda}{2}\int\frac{\diff^3 k}{(2\pi)^3}\frac{1+2 n(\omega)}{2 \omega(\bm{k})}\bigg(1+\lambda\int_{-\infty}^{t}\diff t'\int\frac{\diff^3 q}{(2\pi)^3}\Im\Big[G_{\phi}^{++}(\bm{q},t-t')\Big]^2 \bigg)\\
&+\frac{c_6}{8}\bigg[ \int \frac{d^3 q}{(2 \pi)^3}\frac{1+2 n(\omega)}{2 \omega(\bm q)}\bigg]^2\Bigg] \varphi_c(x)+\bigg[\frac{\lambda}{3!}+\frac{c_6}{12}\int \frac{\diff^3 q}{(2\pi)^3} \frac{1+2 n(\omega)}{2 \omega(\bm{q})}\bigg]\varphi_c^3(x)\\
&+\frac{c_6}{5!}\varphi_c^5(x)+\frac{\lambda^2}{2}\varphi_c(x)\int\diff^3 x'\int_{-\infty}^{t}\diff t' \varphi_c^2(\bm{x}',t')\Im \Big[G_{\phi}^{++}\Big]^2_{x,x'}\\
&+\frac{\lambda^2}{3}\int \diff x' \int_{-\infty}^{t}\diff t' \varphi_c(\bm{x}',t')\Im\Big[G_{\phi}^{++}\Big]^3_{x,x'}=\varphi_c(x)\xi_1(x)+\xi_2(x)\;.
    \end{aligned}
\end{equation}
where $\Big[G_{\phi}^{++}\Big]^2_{x,x'}$ and $\Big[G_{\phi}^{++}\Big]^3_{x,x'}$ are given by Eqs.~(\ref{eq:compact notation1}) and (\ref{eq:compact notation2}), respectively. In order to obtain a Langevin-like equation, we perform a series of approximations to the above equation of motion. We first focus on the last two terms on the left hand side (LHS) of Eq.~\eqref{eq:equation of motion}. Considering Eq.~\eqref{eq:compact notation1}, spatial nonlocality can be effectively addressed by considering only the contributions from zero external momentum. This is a standard procedure in the computation of linear response functions~\cite{Jeon:1992kk,Gleiser:1993ea}, and it is typically employed in the calculation of the one-loop effective potential as an expansion of vertex functions at zero external momentum. Physically, this corresponds to the consideration of nearly spatially uniform field configurations. Consequently, we can get the following expression
\begin{align}\label{eq:patially uniform field approximation}
\frac{\lambda^2}{2}\varphi_c(x)\int\diff^3 x'&\int_{-\infty}^{t}\diff t' \varphi_c^2(\bm{x}',t')\Im \Big[G_{\phi}^{++}\Big]^2_{x,x'}\nn\\
=&\frac{\lambda^2}{2}\varphi_c(\bm{x},t)\int_{-\infty}^{t}\diff t'\Big[\varphi_c^2(\bm{x}',t')-\varphi_c^2(\bm{x},t)\Big]\int \frac{\diff^3 q}{(2\pi)^3}\Im \Big[G_{\phi}^{++}(\bm{q},t-t')\Big]^2\nn\\
&+\frac{\lambda^2}{2}\varphi_c^3(\bm{x},t)\int_{-\infty}^{t}\diff t'\int \frac{\diff^3 q}{(2\pi)^3}\Im \Big[G_{\phi}^{++}(\bm{q},t-t')\Big]^2\;.
\end{align}
To handle the temporal nonlocality, we further assume that the quantity $\varphi_c$ varies sufficiently slowly in time, i.e.,
\begin{equation}
    \varphi^{n}(x')=\varphi^n(x)+n \varphi^{n-1}(x)\dot{\varphi}(x)(t'-t)+\cdots\;.
\end{equation}
This allows us to expand the first term on the right hand side (RHS) of Eq.~\eqref{eq:patially uniform field approximation} to first order around $t$. The assumption is justified for systems near equilibrium, where $\varphi_c$ is not expected to change considerably over time. This approximation has been termed the quasiadiabatic approximation in Ref.~\cite{Morikawa:1986rp}. We then obtain
\begin{align}\label{eq:quasiadiabatic approximation}
\frac{\lambda^2}{2}\varphi_c(x)\int\diff^3 x'&\int_{-\infty}^{t}\diff t' \varphi_c^2(\bm{x}',t')\Im \Big[G_{\phi}^{++}\Big]^2_{x,x'}\nn\\
\simeq&\lambda^2\varphi_c^2(\bm{x},t)\dot{\varphi}_c(\bm{x},t)\int_{-\infty}^{t}(t'-t)\int\frac{\diff^3 q}{(2\pi)^3}\Im \Big[G_{\phi}^{++}(\bm{q},t-t')\Big]^2\nn\\
&+\frac{\lambda^2}{2}\varphi_c^3(\bm{x},t)\int_{-\infty}^{t}\diff t'\int \frac{\diff^3 q}{(2\pi)^3}\Im \Big[G_{\phi}^{++}(\bm{q},t-t')\Big]^2\;.
\end{align}
The last term on the LHS of Eq.~\eqref{eq:equation of motion} can be evaluated in a manner analogous to that shown in Eqs.~\eqref{eq:patially uniform field approximation} and~\eqref{eq:quasiadiabatic approximation}, which yields
\begin{align}\label{eq:quasiadiabatic approximation2}
\frac{\lambda^2}{3}\int \diff x' &\int_{-\infty}^{t}\diff t' \varphi_c(\bm{x}',t')\Im\Big[G_{\phi}^{++}\Big]^3_{x,x'}\nn\\
\simeq&\frac{\lambda^2}{3}\dot{\varphi}_c(\vec{x},t) \int_{-\infty}^{t}\diff t'(t'-t)\Im\Big[ \prod_{j=1}^{3} \int \frac{\diff^3 q_j}{(2\pi)^3}G_{\phi}^{++}(\bm{q}_j,t-t')\Big]\delta(\bm{q}_1+\bm{q}_2+\bm{q}_3)\nn\\
&+\frac{\lambda^2}{3}\varphi_c(\bm{x},t)\int_{-\infty}^{t}\diff t'\Im\Big[ \prod_{j=1}^{3} \int \frac{\diff^3 q_j}{(2\pi)^3}G_{\phi}^{++}(\bm{q}_j,t-t')\Big] \delta(\bm{q}_1+\bm{q}_2+\bm{q}_3)\;.
\end{align}

The first term in Eq.~\eqref{eq:quasiadiabatic approximation2} involves $\dot{\varphi}_c$ and contributes to dissipation coefficient, while the second term gives the two-loop thermal mass correction from the fourth graph in Eq.~\eqref{eq:action picture}. Similarly, the expression in Eq.~\eqref{eq:quasiadiabatic approximation} contributes to the corresponding dissipative term and the thermal correction to the quartic coupling. However, if the free propagator is used in the nonlocal kernels, which leads to vanishing dissipation coefficients. To obtain nonvanishing dissipation coefficients, we need to account for the finite thermal width $\Gamma$ generated by self-energy corrections.

Following Refs.~\cite{Gleiser:1993ea,Parwani:1991gq}, we adopt the dressed propagator scheme by adding and subtracting the self-energy $\Sigma$ in the perturbative expansion. The resummed contribution is absorbed into the propagator, while the subtracted term is retained in the interaction to avoid double counting. The dressed propagator is characterized by a temperature-dependent quasiparticle dispersion relation and a finite thermal width $\Gamma$,
\begin{equation}\label{eq:the dressed propagator}
    G^{++}_{\phi}(\bm{q},t-t')=\frac{e^{-\Gamma|t-t'|}}{2 \omega(\bm{q})}\Big[ (1+2n)\cos[\omega(t-t')]+2\beta\Gamma n(1+n) \sin[\omega |t-t'|]-i\sin[\omega|t-t'|]\Big]\,.
\end{equation}
The real and imaginary parts of the self-energy determines respectively the thermal mass $m_T$ and decay width $\Gamma$~\cite{Landsman:1986uw},
\begin{equation}
     m_T^2=m^2+\Re \Sigma \;,\qquad \Gamma(q)=-\frac{\Im \Sigma(q)}{2 \omega(q)}\;.\label{eq:Particle decay width definition}
\end{equation}

The finite width is essential for the dissipative dynamics. The leading contribution to the imaginary part of the self-energy from the setting sun diagram in the fourth graph in Eq.~\eqref{eq:action picture}. The sextic interaction contributes only to the real part of the self-energy at this order. Therefore, the leading-order damping width is the same as that in the $\lambda \phi^4$ theory~\cite{Gleiser:1993ea}. In the narrow-width approximation, $\beta \Gamma\ll1$, where $\beta=1/T$, the decay width in Eq.~\eqref{eq:Particle decay width definition}  in the high temperature limit is given in Refs.~\cite{Hosoya:1983ke,Parwani:1991gq}, then $\Gamma$ is
\begin{equation}\label{eq:Gamma result}
	\Gamma \simeq \frac{\lambda^2 T^2}{1536 \pi \omega}\;.
\end{equation}
The dressed propagator is characterized by the temperature dependent quasiparticle energy $\omega(\bm{q})=\sqrt{\bm{q}^2+m_T^2}$. The thermal mass and quartic coupling include the corresponding local thermal correction in Eq.~\eqref{eq:equation of motion}. The thermal mass receives contributions from the two-point diagrams, and the thermal quartic coupling is calculated from the four-point diagrams.
In the high temperature limit, the finite temperature mass $m_T$ is
\begin{equation}\label{eq:finite temperature mass}
    m_T^2=m^2+\frac{\lambda T^2}{24}-\frac{\lambda mT}{8\pi}-\frac{\lambda^2 T^3}{384 \pi m}+\frac{\lambda^2 T^2}{128 \pi^2}+\frac{\lambda^2 T^2}{192 \pi^2}\ln \frac{\bar{\mu}^2}{T^2}+\frac{c_6T^4}{1152}-\frac{c_6 mT^3}{192 \pi}+\frac{c_6 m^2 T^2}{128 \pi^2}+\cdots\;,
\end{equation}
and $\lambda_T$ is given by
\begin{equation}\label{eq:lambdaT}
    \lambda_T=\lambda-\frac{3\lambda^2 T}{16\pi m}-\frac{3\lambda^2}{(4\pi)^2}\ln\lp \frac{\bar{\mu}e^{\gammaE}}{4\pi T}\rp+\frac{c_6 T^2}{24}-\frac{c_6 mT}{8\pi}-\frac{c_6 m^2}{(4\pi)^2}\ln\lp \frac{\bar{\mu}e^{\gammaE}}{4\pi T}\rp\;.
\end{equation}
Substituting the dressed propagator in Eq.~\eqref{eq:the dressed propagator} into the dissipation term in the first term on the RHS of Eq.~\eqref{eq:quasiadiabatic approximation}, we can obtain the local dissipation term, $\eta_1\varphi_c^2\dot{\varphi}_c$, where dissipation coefficient $\eta_1$ is given by
\begin{equation}\label{eq:dissipation coefficient eta1}
    \eta_1=\frac{\lambda^2}{8}\beta\int \frac{\diff^3 q}{(2\pi)^3}\frac{n(1+n)}{\omega^2 \Gamma}+\mathcal{O}(\lambda^2\frac{\Gamma}{\omega})\;.
\end{equation}
Using Eq.~\eqref{eq:Gamma result} in Eq.~\eqref{eq:dissipation coefficient eta1}, we obtain the $\eta_1$ in the high temperature limit,
\begin{equation}\label{eq:dissipation term eta1}
    \eta_1=\frac{192 \pi}{T^3}\int\frac{\diff^3 q}{(2 \pi)^3}\frac{n(1+n)}{\omega}\simeq \frac{96}{\pi T}\ln\lp \frac{T}{m_T}\rp \;.
\end{equation}
We can proceed in an analogous way and evaluate Eq.~\eqref{eq:quasiadiabatic approximation2} in order to obtain the expression for the dissipation coefficient associated with the second fluctuation (noise) field $\xi_2$, but the corresponding dissipation coefficient associated with $\xi_2$ is at least of the order $O(\lambda^4)$. Therefore, in a weakly interacting model, the dominant contribution to dissipation in the equation of motion for $\varphi_c$ comes from the dissipation term associated with the multiplicative noise field $\xi_1$.

At last, we obtain the final Langevin equation
\begin{equation}
\Big[\Box +m_T^2\Big]\varphi_c(\bm{x},t)+\frac{\lambda_T}{3!}\varphi^3_c(\bm{x},t)+\frac{c_6}{5!}\varphi_c(\bm{x},t)^5+\eta_1\varphi^2_c(\bm{x},t)\dot{\varphi}_c(\bm{x},t)=\varphi_c(\bm{x},t) \xi_1(\bm{x},t)\;,
\end{equation}
where $m_T$ is given by Eq.~\eqref{eq:finite temperature mass}, $\lambda_T$ is given by Eq.~\eqref{eq:lambdaT} and $\eta_1$ is the dissipation term given in Eq.~\eqref{eq:dissipation term eta1} and $\xi_1$ is the fluctuation term satisfies Gaussian distributions given by Eq.~\eqref{eq:probability distribution1}. The two-point correlation function for $\xi_1$ reads 
\begin{equation}
    \braket{\xi_1(x)\xi_1(x')}=\frac{\lambda^2}{2}\Re\Big[G^{++}_{\phi}\Big]_{x,x'}\;.
\end{equation}
In the Markovian limit, that the correlation time of the thermal fluctuations is much shorter than the characteristic time scale of the background-field evolution, the nonlocal noise becomes local. We then obtain the fluctuation-dissipation relation,
\begin{equation}
    \braket{\xi_1(x)\xi_1(x')}=2T\eta_1\delta(\bm x-\bm{x}')\delta(t-t')\;.
\end{equation}

\section{The two field model in the CTP}\label{sec:two field model}
We now consider a system comprising two interacting real singlet scalar fields, $H$ and $S$. This system often serves as a paradigm for the scalar sector in electroweak PTs or Higgs‑portal scalar dark matter models. The Lagrangian density of the two-field model is given by
\begin{equation}
    \mathcal{L}[H,S]=\frac{1}{2}(\partial_\mu H)^2+\frac{1}{2}(\partial_\mu S)^2-V[H,S]\,,
\end{equation}
where
\begin{equation}
V[H,S]=\frac{1}{2}m_h^2 H^2+\frac{\lambda_h}{4} H^4+\frac{1}{2}m_s^2 S^2+\frac{\lambda_s}{4} S^4+\frac{\lambda_{hs}}{4}H^2S^2\;.
\end{equation}
To facilitate the background-field decomposition, we denote the tree-level classical field values corresponding to these two fields as $\phi$ and $\chi$, respectively. Substituting the quantum fields with their classical backgrounds, the tree-level effective potential reads
\begin{equation}
	V(\phi,\chi)=\frac{1}{2}m_h^2 \phi^2+\frac{\lambda_h}{4} \phi^4+\frac{1}{2}m_s^2 \chi^2+\frac{\lambda_s}{4} \chi^4+\frac{\lambda_{hs}}{4}\phi^2\chi^2\;.
\end{equation}

To investigate the nonequilibrium dynamics of this system within the quantum field theory framework, we introduce external sources $J_H(x)$ and $J_S(x)$, which are conjugate to the fields $H$ and $S$, the corresponding generating functional $Z[J_H,J_S]$ is defined via the path integral:
\begin{equation}
    Z[J_H,J_S]=\int\mathcal{D}H\mathcal{D}S \exp\Big\{iS[H,S,J_H,J_S]\Big\}\;,
\end{equation}
where the classical action $S[H,S,J_H,J_S]$ is given by the integration of the Lagrangian density and source terms over the CTP contour,
\begin{equation}
S[H,S,J_H,J_S]=\int_c \diff^4 x \Big\{\mathcal{L}[H,S]+J_H(x)H(x)+J_S(x)S(x) \Big\}\,.
\end{equation}
We next define the connected generating functional $W[J_H,J_S]$ as
\begin{equation}
    W[J_H,J_S]=-i \ln Z[J_H,J_S]\;.
\end{equation}
By taking the functional derivatives with respect to the sources, we obtain the classical background fields
\begin{equation}
	\phi(x)\equiv\frac{\delta W[J_H,J_S]}{\delta J_H(x)}\;, \quad \chi(x)\equiv\frac{\delta W[J_H,J_S]}{\delta J_S(x)}\;.
\end{equation}
Subsequently, the effective action $\Gamma[\phi,\chi]$ is defined as the Legendre transform of $W[J_H,J_S]$ with respect to the external sources,
\begin{equation}
\Gamma[\phi,\chi]=W[J_H,J_S]-\int_c \diff^4 x \Big(J_H(x)\phi(x)+J_S(x)\chi(x)\Big)\;.
\end{equation}
With this definition, one immediately obtains the variational relations of the effective action with respect to the classical fields:
\begin{equation}
	\frac{\delta \Gamma}{\delta \phi(x)}=-J_H(x)\;,\quad \frac{\delta \Gamma}{\delta \chi(x)}=-J_S(x)\;.
\end{equation}
When the external sources vanish ($J_H=J_S=0$), the extremum condition of the effective action determines the vacuum structure of the system.

To explicitly evaluate the effective action, we decompose the quantum fields into classical backgrounds and their corresponding quantum fluctuations,
\begin{equation}
	H(x)=\phi(x)+h(x)\;,\quad S(x)=\chi(x)+s(x)\;.
\end{equation}
Here, $\phi$ and $\chi$ denote the classical background fields, while $h$ and $s$ are the quantum fluctuations. By expanding the potential, the quadratic terms of the fluctuations yield a mass matrix that mixes the $h$ and $s$ fields. 
The quadratic fluctuation term can be written as
\begin{equation}
    V^{(2)}=\frac{1}{2}\begin{pmatrix}
        h & s
    \end{pmatrix}\mathcal{M}^2(\phi,\chi)\begin{pmatrix}
        h\\s
    \end{pmatrix}\;,
\end{equation}
where the background dependent mass matrix is
\begin{equation}\label{eq:mass matrix}
    \mathcal{M}^2=\begin{pmatrix}
        m_h^2+3\lambda_h \phi^2+\frac{\lambda_{hs}}{2}\chi^2&\lambda_{hs}\phi\chi\\
        \lambda_{hs}\phi\chi& m_s^2+3\lambda_s \chi^2+\frac{\lambda_{hs}}{2}\phi^2
    \end{pmatrix}.
\end{equation}
The Higgs-portal interaction therefore induces a background-dependent mixing between the two fluctuation fields. Accordingly, the full CTP propagator in the original field basis of $(h,s)$ compose the matrix of,
\begin{equation}
    G^{ab}(x,x')=\begin{pmatrix}
    G^{ab}_{hh}(x,x')&G^{ab}_{hs}(x,x')\\
    G^{ab}_{hs}(x,x')&G^{ab}_{ss}(x,x')
    \end{pmatrix},\quad a,b=\pm,
\end{equation}
and its off-diagonal components need not vanish when both background fields are nonzero.
In the following, we retain the original field basis rather than performing a background-dependent diagonalization of $\mathcal{M}^2$ in Eq.~\eqref{eq:mass matrix}. This is particularly convenient for nonequilibrium configurations, for which the background fields are generally spacetime dependent and a local diagonalization would introduce additional derivative couplings through the spacetime dependence of the rotation matrix.

For the perturbative calculation of the effective action, we separate the quadratic fluctuation operator into a background independent part and a background dependent interaction part.

\begin{figure}[!tbp]
	\centering
	\includegraphics[width=0.9\linewidth]{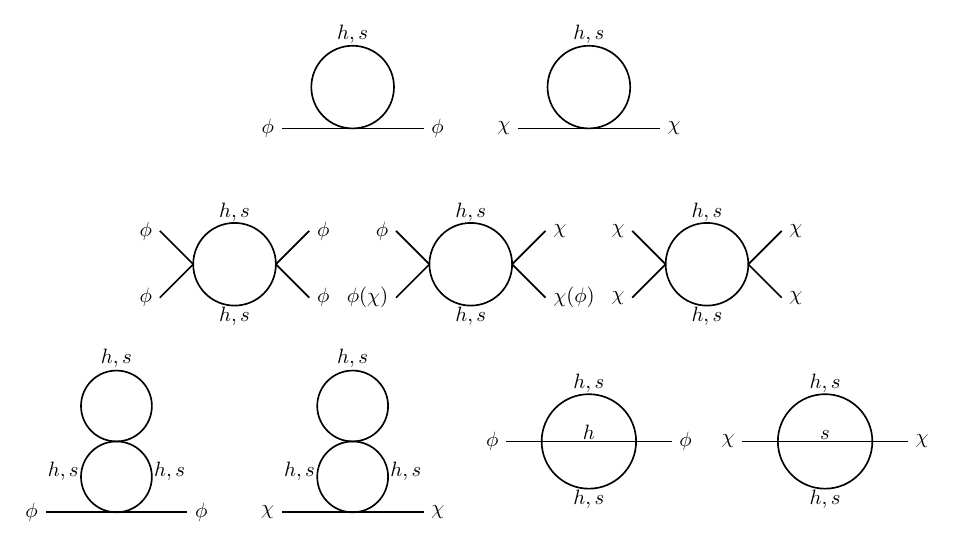}
	\caption{Connected Feynman topologies of the two-field model at the two-loop order. The external solid lines correspond to the background fields $\phi$ and $\chi$, while the internal loop lines denote the thermal propagators for $h$ and $s$.}
	\label{fig:double-field-action}
\end{figure}

Truncating our perturbative expansion at the two-loop order and up to $\lambda^2$, the relevant connected Feynman diagrams contributing to the effective potential are depicted in Figure~\ref{fig:double-field-action}.
The detailed analytic evaluations of the topologies in Figure~\ref{fig:double-field-action} are presented in Appendix~\ref{appendix:two-field model}.

Building on these results, the CTP effective action is therefore written as
\begin{equation}
    \Gamma_{\rm CTP}=S[\phi_+,\chi_+]-S[\phi_-,\chi_-]+\Gamma_{\rm loop}[\phi_{\pm},\chi_{\pm}]\;.
\end{equation}
To derive the Langevin equation from the effective action, we introduce
\begin{equation}
    \phi_c=\frac{\phi_++\phi_-}{2}, \quad \phi_\Delta=\phi_+-\phi_-,
\end{equation}
and similarly
\begin{equation}
    \chi_c=\frac{\chi_++\chi_-}{2}, \quad \chi_\Delta=\chi_+-\chi_-\;.
\end{equation}
After transforming the loop contributions to the physical-response basis~\cite{Gleiser:1993ea,Bellac:1996Le}, the effective action can be organized in the generic form
\begin{align}
\Gamma_{\rm CTP}=&\int \diff^4x\Bigg\{\phi_\Delta(-\Box-m_h^2)\phi_c+ \chi_\Delta(-\Box-m_s^2)\chi_c-\frac{\lambda_h}{4}\big[4\phi_\Delta\phi_c^3+\phi_\Delta^3\phi_c\big]-\frac{\lambda_s}{4}\big[4\chi_\Delta\chi_c^3+\chi_\Delta^3\chi_c\big]\nn \\
&-\frac{\lambda_{hs}}{4}\big[ \frac{1}{2}\phi_\Delta^2\chi_\Delta\chi_c+\frac{1}{2}\phi_\Delta\chi_\Delta^2\phi_c+2\phi_\Delta\phi_c\chi_c^2+2\chi_\Delta\phi_c^2\chi_c\big]+\phi_\Delta\big[-3\lambda_h I(m_h)-\frac{1}{2}\lambda_{hs}I(m_s) \big]\phi_c\nn\\
&+\chi_\Delta\big[-3\lambda_s I(m_s)-\frac{1}{2}\lambda_{hs}I(m_h) \big]\chi_c-\phi_\Delta\big[18\lambda_h^2\mathcal{I}_3(m_h,m_h,m_h)+\frac{\lambda_{hs}^2}{2}\mathcal{I}_3(m_s,m_s,m_h)\nn\\
&+3\lambda_h\lambda_{hs}\mathcal{I}_3(m_h,m_h,m_s)+3\lambda_s\lambda_{hs}\mathcal{I}_3(m_s,m_s,m_s)\big]\phi_c-\chi_\Delta\big[ 18\lambda_s^2\mathcal{I}_3(m_s,m_s,m_s)\nn\\
&+\frac{\lambda_{hs}^2}{2}\mathcal{I}_3(m_h,m_h,m_s)+3\lambda_h\lambda_{hs}\mathcal{I}_3(m_h,m_h,m_h)+3\lambda_s\lambda_{hs}\mathcal{I}_3(m_s,m_s,m_h)\big]\chi_c\Bigg\}\nn\\
&+\int\diff^4x\int\diff^4x'\Bigg\{-\bigg[\frac{9}{2}\lambda_h^2\Im\Big[G^{++}_{h,h}\Big]^2+\frac{1}{8}\lambda_{hs}^2\Im\Big[G^{++}_{s,s}\Big]^2\bigg]\Big[4\phi_c^2(x)\phi_\Delta(x')\phi_c(x')\nn\\
&+\phi_\Delta^2(x)\phi_\Delta(x')\phi_c(x')\Big]-\frac{3}{4}\lp\lambda_h\lambda_{hs}\Im\Big[G^{++}_{h,h}\Big]^2+\lambda_s\lambda_{hs}\Im\Big[G^{++}_{s,s}\Big]^2\rp\bigg[\phi_c(x')\phi_\Delta(x')\nn\\
&\times \Big(4\chi^2(x)+\chi_\Delta^2(x)\Big)+ \Big(4\phi_c^2(x)+\phi_\Delta^2(x)\Big) \chi_c(x')\chi_\Delta(x')\bigg]-\frac{\lambda_{hs}^2}{2}\Im\Big[G^{++}_{h,s}\Big]^2\Big(4\phi_c(x)\chi_c(x)\nn\\
&+\phi_\Delta(x)\chi_\Delta(x)\Big)\times \Big(\phi_\Delta(x')\chi_c(x')+\phi_c(x')\chi_\Delta(x')\Big)-\Big[\frac{9}{2}\lambda_s^2\Im\Big[G^{++}_{s,s}\Big]^2+\frac{1}{8}\lambda_{hs}^2\Im\Big[G^{++}_{h,h}\Big]^2 \Big]\nn\\
&\times [4\chi_c^2(x)\chi_\Delta(x')\chi_c(x')+\chi_\Delta^2(x)\chi_\Delta(x')\chi_c(x')]-\bigg[12\lambda_h^2\Im\Big[G^{++}_{h,h,h}\Big]^3+\lambda_{hs}^2\Im\Big[G^{++}_{h,s,s}\Big]^3\bigg]\nn\\
&\times \phi_\Delta(x)\phi_c(x')-\bigg[12\lambda_s^2\Im\Big[G^{++}_{s,s,s}\Big]^3+\lambda_{hs}^2\Im\Big[G^{++}_{h,h,s}\Big]^3\bigg]\chi_\Delta(x)\chi_c(x')\Bigg\}\theta(t-t')\nn\\
&+i\int\diff^4x\int\diff^4x'\Bigg\{\phi_\Delta(x)\phi_c(x)\phi_\Delta(x')\phi_c(x')\bigg[ 9\lambda_h^2\Re\Big[G^{++}_{h,h}\Big]^2+\frac{\lambda_{hs}^2}{4}\Re\Big[G^{++}_{s,s}\Big]^2\bigg]\nn\\
&+\frac{\lambda_{hs}^2}{2}\Re\Big[G^{++}_{h,s} \Big]^2\bigg[\phi_\Delta(x)\chi_c(x)+\chi_\Delta(x)\phi_c(x)\bigg]\times\bigg[\phi_\Delta(x')\chi_c(x')+\chi_\Delta(x')\phi_c(x') \bigg]\nn\\
&+\phi_\Delta(x)\phi_c(x)\chi_\Delta(x')\chi_c(x')\bigg[ 3\lambda_h\lambda_{hs}\Re\Big[G^{++}_{h,h}\Big]^2+3\lambda_s\lambda_{hs}\Re\Big[G^{++}_{s,s}\Big]^2\bigg]\nn\\
&+\chi_\Delta(x)\chi_c(x)\chi_\Delta(x')\chi_c(x')\bigg[ 9\lambda_s^2\Re\Big[G^{++}_{s,s}\Big]^2+\frac{\lambda_{hs}^2}{4}\Re\Big[G^{++}_{h,h}\Big]^2\bigg]\nn\\
&+\phi_\Delta(x)\phi_c(x')\bigg[3\lambda_h^2\Re\Big[G^{++}_{h,h,h}\Big]^3+\frac{\lambda_{hs}^2}{4}\Re\Big[G^{++}_{h,s,s}\Big]^3\bigg]\nn\\
&+ \chi_\Delta(x)\chi_c(x')\bigg[3\lambda_s^2\Re\Big[G^{++}_{s,s,s}\Big]^3+\frac{\lambda_{hs}^2}{4}\Re\Big[G^{++}_{h,h,s}\Big]^3\bigg]
\Bigg\}\;,
\end{align}
where 
\begin{equation}
    \Big[G^{++}_{{\Psi_1,\cdots,\Psi_n}}\Big]^n=\int\frac{\diff^3 k}{(2\pi)^3}\exp[i\bm{k}\cdot(\bm{x}-\bm{x}')]\bigg[\prod_{j=1}^n\int\frac{\diff^3 q_j}{(2\pi)^3}G^{++}_{\Psi_j}(\bm{q}_j,t-t')\bigg]\delta(\bm{k}-\sum_j \bm{q}_j).
\end{equation}
Having obtained the CTP effective action in the physical-response basis, we now derive the corresponding nonequilibrium equations of motion. The physical dynamics follows from varying the effective action with respect to the difference fields and subsequently taking the physical limit,
\begin{equation}
    \left.\frac{\partial \Gamma_{\rm CTP}}{\partial \phi_\Delta(x)}\right|_{\phi_\Delta=\chi_\Delta=0}=0\,,\qquad \left.\frac{\partial \Gamma_{\rm CTP}}{\partial \chi_\Delta(x)}\right|_{\phi_\Delta=\chi_\Delta=0}=0\;.
\end{equation}
In this limit, the physical fields are identified with the background fields,
\begin{equation}
    \phi_c\to\phi,\qquad \chi_c\to\chi\;.
\end{equation}
Therefore, we can obtain
\begin{align}
(\Box+&m_h^2)\phi(x)+\frac{\lambda_{h}}{4}4\phi^3(x)+\frac{\lambda_{hs}}{4}2\phi(x)\chi^2(x)+\Big[3\lambda_h I(m_h)+\frac{\lambda_{hs}}{2}I(m_s)\Big]\phi(x)\nn\\
&+\Big[18\lambda_h^2\mathcal{I}_3(m_h,m_h,m_h)+\frac{\lambda_{hs}^2}{2}\mathcal{I}_3(m_s,m_s,m_h)+3\lambda_h\lambda_{hs}\mathcal{I}_3(m_h,m_h,m_s)\nn\\
&+3\lambda_s\lambda_{hs}\mathcal{I}_3(m_s,m_s,m_s)\Big]\phi(x)\nn+\phi(x)\int_{x'}\bigg[\Big(18\lambda_{h}^2 \Im\Big[G^{++}_{h,h}\Big]^2+\frac{\lambda_{hs}^2}{2}\Im\Big[G^{++}_{s,s}\Big]^2\Big)\phi^2(x')\nn\\
&+\Big(3\lambda_{h}\lambda_{hs}\Im\Big[G^{++}_{h,h}\Big]^2+3\lambda_{hs}\lambda_s\Im\Big[G^{++}_{s,s}\Big]^2\Big)\chi^2(x')\bigg]+\chi(x)\int_{x'}2\lambda_{hs}^2\Im\Big[G^{++}_{h,s}\Big]^2\phi(x')\chi(x')\nn\\
&+\int_{x'} \Big[12\lambda_h^2\Im\Big[G^{++}_{h,h,h}\Big]^3+\lambda_{hs}^2 \Im\Big[G^{++}_{h,s,s}\Big]^3\Big]\phi(x')\nn\\
&=\phi(x)\xi_{h,1}+\chi(x)\xi_{h,2}+\xi_{h,3}\;,
\end{align}

and

\begin{align}
(\Box+&m_s^2)\chi(x)+\frac{\lambda_{s}}{4}4\chi^3(x)+\frac{\lambda_{hs}}{4}2\chi(x)\phi^2(x)+\Big[3\lambda_s I(m_s)+\frac{\lambda_{hs}}{2}I(m_h)\Big]\chi(x)\nn\\
&+\big[18\lambda_s^2\mathcal{I}_3(m_s,m_s,m_s)+\frac{\lambda_{hs}^2}{2}\mathcal{I}_3(m_h,m_h,m_s)+3\lambda_h\lambda_{hs}\mathcal{I}_3(m_h,m_h,m_h)\nn\\
&+3\lambda_s\lambda_{hs}\mathcal{I}_3(m_s,m_s,m_h)\big]\chi(x)+\chi(x)\int_{x'}\bigg[\Big(18\lambda_{s}^2 \Im\Big[G^{++}_{s,s}\Big]^2+\frac{\lambda_{hs}^2}{2}\Im\Big[G^{++}_{h,h}\Big]^2\Big)\chi^2(x')\nn\\
&+\Big(3\lambda_{s}\lambda_{hs}\Im\Big[G^{++}_{s,s}\Big]^2+3\lambda_{hs}\lambda_h\Im\Big[G^{++}_{h,h}\Big]^2\Big)\phi^2(x')\bigg]+\phi(x)\int_{x'}2\lambda_{hs}^2\Im\Big[G^{++}_{h,s}\Big]^2\phi(x')\chi(x')\nn\\
&+\int_{x'} \Big[12\lambda_s^2\Im\Big[G^{++}_{s,s,s}\Big]^3+\lambda_{hs}^2 \Im\Big[G^{++}_{h,h,s}\Big]^3\Big]\chi(x')\nn\\
&=\chi(x)\xi_{s,1}+\phi(x)\xi_{s,2}+\xi_{s,3}\;.
\end{align}

Where $\xi_{i,a} (\text{for i=h,s. \text{and} a=1,2,3})$ are Gaussian fluctuation fields, which satisfy
\begin{align}\label{eq:relation1}
	\braket{\xi_{h,1}(x)\xi_{h,1}(x')}&=18\lambda_h^2\Re\Big[G^{++}_{h,h}\Big]^2+\frac{\lambda_{hs}^2}{2}\Re\Big[G^{++}_{s,s}\Big]^2\;,\\
	\braket{\xi_{h,2}(x)\xi_{h,2}(x')}&=\lambda_{hs}^2\Re\Big[G^{++}_{h,s} \Big]^2\;,\\
	\braket{\xi_{h,3}(x)\xi_{h,3}(x')}&=6\lambda_h^2\Re\Big[G^{++}_{h,h,h}\Big]^3+\frac{\lambda_{hs}^2}{2}\Re\Big[G^{++}_{h,s,s}\Big]^3\;,\\
    \braket{\xi_{s,1}(x)\xi_{s,1}(x')}&=18\lambda_s^2\Re\Big[G^{++}_{s,s}\Big]^2+\frac{\lambda_{hs}^2}{2}\Re\Big[G^{++}_{h,h}\Big]^2\;,\\
	\braket{\xi_{s,2}(x)\xi_{s,2}(x')}&=\lambda_{hs}^2\Re\Big[G^{++}_{h,s} \Big]^2\;,\\
	\braket{\xi_{s,3}(x)\xi_{s,3}(x')}&=6\lambda_s^2\Re\Big[G^{++}_{s,s,s}\Big]^3+\frac{\lambda_{hs}^2}{2}\Re\Big[G^{++}_{h,h,s}\Big]^3\;,\\
    \braket{\xi_{h,1}(x)\xi_{s,1}(x')}&=3\lambda_h\lambda_{hs}\Re\Big[G^{++}_{h,h}\Big]^2+3\lambda_s\lambda_{hs}\Re\Big[G^{++}_{s,s}\Big]^2\;,\\
    \braket{\xi_{h,2}(x)\xi_{s,2}(x')}&=\lambda_{hs}^2\Re\Big[G^{++}_{h,s} \Big]^2\;.\label{eq:relation7}
\end{align}
The last two terms represent the coupling effect arising from the two fields, while the remaining terms are mutually independent. As in the single-field treatment aforementioned, we retain the leading contributions and further obtain the coupled Langevin equations: 
\begin{align}\label{eq:double L1}
    \Box \phi(x)+\frac{\partial V_{\rm eff}(\phi,\chi)}{\partial \phi(x)}+\eta_{\phi\phi}\dot{\phi}(x)+\eta_{\phi\chi}\dot{\chi}(x)&=\phi(x)\xi_{h,1}+\chi(x)\xi_{h,2}\;,\\ 
    \Box \chi(x)+\frac{\partial V_{\rm eff}(\phi,\chi)}{\partial \chi(x)}+\eta_{\chi\phi}\dot{\phi}(x)+\eta_{\chi\chi}\dot{\chi}(x)&=\chi(x)\xi_{s,1}+\phi(x)\xi_{s,2}\;, \label{eq:double L2}
\end{align}
where $V_{\rm eff}$ denote the thermal effective potential, that reads
\begin{equation}
    V_{\rm eff}(\phi,\chi)=\frac{1}{2}m_{h,T}^2\phi^2+\frac{\lambda_{h,T}}{4}\phi^4+\frac{1}{2}m_{s,T}^2\chi^2+\frac{\lambda_{s,T}}{4}\chi^4+\frac{\lambda_{hs,T}}{4}\phi^2\chi^2\;,
\end{equation}
with
\begin{align}
    m_{h,T}^2=&m_h^2+T^2 \left(\frac{9 \lambda_h^2}{32 \pi ^2}+\frac{\lambda_h}{4}+\frac{3 \lambda_{hs} \lambda_s}{64 \pi ^2}+\frac{\lambda_{hs}}{24}+\frac{3 \lambda_h \lambda_{hs} m_s}{64 \pi ^2 m_h}+\frac{\lambda_{hs}^2 m_h}{128 \pi ^2 m_s}\right)\nn\\
    &+T^3 \left(-\frac{3 \lambda_h^2}{32 \pi  m_h}-\frac{\lambda_h \lambda_{hs}}{64 \pi  m_h}-\frac{\lambda_{hs}^2}{384 \pi  m_s}-\frac{\lambda_{hs} \lambda_s}{64 \pi  m_s}\right)+T \left(-\frac{3 \lambda_h m_h}{4 \pi }-\frac{\lambda_{hs} m_s}{8 \pi }\right)\;,\nn\\
    \lambda_{h,T}=&\lambda_h-\frac{9\lambda_h^2 T}{8\pi m_h}-\frac{9\lambda_{h}^2}{8\pi^2}\ln\lp \frac{\bar{\mu}e^{\gammaE}}{4\pi T}\rp-\frac{\lambda_{hs}^2T}{32\pi m_s}-\frac{\lambda_{hs}^2}{32\pi^2}\ln\lp \frac{\bar{\mu}e^{\gammaE}}{4\pi T}\rp\;,\nn\\
    \lambda_{hs,T}=&\lambda_{hs}-\frac{3\lambda_h \lambda_{hs}T}{8\pi m_h}-\frac{\lambda_{hs}^2T}{2\pi(m_h+m_s)}-\frac{3\lambda_s \lambda_{hs}T}{8\pi m_s}-\frac{\lambda_{hs}(3\lambda_h+2\lambda_{hs}+3\lambda_s)}{8\pi^2}\ln\lp \frac{\bar{\mu}e^{\gammaE}}{4\pi T}\rp\;,\nn\\
   m_{s,T}^2=& \left. m_{h,T}^2\right|_{h\leftrightarrow s}\;,\qquad \lambda_{s,T}=\left. \lambda_{h,T}\right|_{h\leftrightarrow s}\;.
\end{align}
Here, $h\leftrightarrow s$ denotes the simultaneous interchange of the corresponding parameters of the two scalar sectors, and $\eta_{ij}(i,j=\phi,\chi)$ denote the dissipation coefficients, which are given by
\begin{align}
\eta_{\phi\phi}&=36\lambda_h^2\phi^2\gamma_{hh}+\lambda_{hs}^2\phi^2\gamma_{ss}+2\lambda_{hs}^2\chi^2\gamma_{hs}\;,\nn\\
\eta_{\chi\chi}&=36\lambda_s^2\chi^2\gamma_{ss}+\lambda_{hs}^2\chi^2\gamma_{hh}+2\lambda_{hs}^2\phi^2\gamma_{hs}\;,\nn\\
\eta_{\phi\chi}=\eta_{\chi\phi}&=6\lambda_h\lambda_{hs}\phi\chi\gamma_{hh}+6\lambda_{hs}\lambda_{s}\phi\chi\gamma_{ss}+2\lambda_{hs}^2 \phi\chi\gamma_{hs}\;,\label{eq:dissipation coefficients}
\end{align}
with
\begin{align}
\gamma_{\Psi_i\Psi_j}=-\int\frac{\diff^3 q}{(2\pi)^3}\int_0^{\infty}\diff\tau\tau\Im\Big[G^{++}_{\Psi_i}(\bm q,\tau)G^{++}_{\Psi_j}(\bm q,\tau)\Big]\;.
\end{align}

To obtain a local dissipative description, we work in the zero-external-momentum and quasiadiabatic limits, assuming that the background fields vary slowly compared with the microscopic relaxation time scale. We further employ the dressed thermal propagators in the narrow-width regime, $\beta\Gamma_i\ll1$ and $\Gamma_i/\omega_i\ll1$, with $i=h,s$. The two independent characteristic integrals have a structure similar to that of the single-field case, whereas the mixed case contains interference contributions between different quasiparticle poles and thus must be treated separately. In practice, the contribution of $\gamma_{hs}$ is negligible since the coupling $\lambda_{hs}$ is negligibly small. Retaining only the leading-order contributions in $1/\Gamma$, the nonlocal characteristic integrals can be approximated as follows,
\begin{equation}\label{eq:characteristic integrals gamma}
    \gamma_{hh} \simeq \beta\int\frac{\diff^3q}{(2\pi)^3}\frac{n_{h}(1+n_h)}{8\omega_h^2\Gamma_h}\;,\qquad
    \gamma_{ss} \simeq \beta\int\frac{\diff^3q}{(2\pi)^3}\frac{n_{s}(1+n_s)}{8\omega_s^2\Gamma_s}\;.
\end{equation}
The expression for the dissipation coefficient can be further simplified if we consider the high temperature limit $T\gg m_{h,T},m_{s,T}$. In this situation, the mass dependence of the phase-space integrals entering the $hhh(sss)$ and $hss(shh)$ sunset contributions is subleading, the thermal widths are therefore given by
\begin{equation}\label{eq: width Gamma for h and s}
    \Gamma_{h}(\bm q)\simeq \frac{T^2}{512 \pi \omega_h(\bm q)}\lp 12\lambda_{h}^2+\lambda_{hs}^2\rp\;,\quad \Gamma_{s}(\bm q)\simeq \frac{T^2}{512 \pi \omega_s(\bm q)}\lp 12\lambda_{s}^2+\lambda_{hs}^2\rp\;.
\end{equation}
Here $\omega_i(\bm q)=\sqrt{\bm{q}^2+m_{i,T}^2}$, with $i=h,s$. Using Eq.~\eqref{eq: width Gamma for h and s} in Eq.~\eqref{eq:characteristic integrals gamma}, we obtain, in the high temperature limit,
\begin{equation}
\gamma_{hh}\simeq\frac{32}{\pi T(12\lambda_{h}^2+\lambda_{hs}^2)}\ln\lp\frac{T}{m_{h,T}}\rp\;,\quad \gamma_{ss}\simeq\frac{32}{\pi T(12\lambda_{s}^2+\lambda_{hs}^2)}\ln\lp\frac{T}{m_{s,T}}\rp\;.
\end{equation}
To find the fluctuation-dissipation relation, we redefine the quantities appearing on the RHS of Eqs.~\eqref{eq:double L1} and~\eqref{eq:double L2} as follows:
\begin{align}
    \Xi_{\phi}=\phi \xi_{h,1}+\chi \xi_{h,2}\;,\quad \Xi_{\chi}=\chi \xi_{s,1}+\phi \xi_{s,2}\;.
\end{align}
According to Eqs.~\eqref{eq:relation1}-\eqref{eq:relation7}, we obtain
\begin{align}
\braket{\Xi_{\phi}(x)\Xi_{\phi}(x')}&=18\lambda_h^2\phi(x)\phi(x')\Re\Big[G^{++}_{h,h}\Big]^2+\frac{\lambda_{hs}^2}{2}\phi(x)\phi(x')\Re\Big[G^{++}_{s,s}\Big]^2+\lambda_{hs}^2\chi(x)\chi(x')\Re\Big[G^{++}_{h,s} \Big]^2\;,\nn\\
\braket{\Xi_{\chi}(x)\Xi_{\chi}(x')}&=18\lambda_s^2\chi(x)\chi(x')\Re\Big[G^{++}_{s,s}\Big]^2+\frac{\lambda_{hs}^2}{2}\chi(x)\chi(x')\Re\Big[G^{++}_{h,h}\Big]^2+\lambda_{hs}^2\phi(x)\phi(x')\Re\Big[G^{++}_{h,s} \Big]^2\;,\nn\\ 
\braket{\Xi_{\phi}(x)\Xi_{\chi}(x')}&=3\lambda_h\lambda_{hs}\phi(x)\chi(x')\Re\Big[G^{++}_{h,h}\Big]^2+ 3\lambda_s\lambda_{hs}\phi(x)\chi(x')\Re\Big[G^{++}_{s,s}\Big]^2+\lambda_{hs}^2\chi(x)\phi(x')\Re\Big[G^{++}_{h,s} \Big]^2\;.
\end{align}
Following the Markovian approximation introduced in the single-field case, keeping the leading contribution in the narrow-width approximation, we have 
\begin{equation}
    \Re\Big[G^{++}_{i,j}\Big]^2\simeq 4 T \gamma_{ij}\delta^{(4)}(x-x')\;,\quad i,j=h,s.
\end{equation}
The stochastic forces are constrained by the same thermal response.
In this limit, the colored Gaussian noises reduce to local white noises, and the noise correlations are related to the dissipative coefficients through the fluctuation–dissipation relation,
\begin{equation}\label{eq:fluctuation–dissipation relation2}
    \braket{\Xi_{i}(x)\Xi_{j}(x')}=2T\eta_{ij}\delta^{(4)}(x-x')\;,\quad i,j=\phi,\chi.
\end{equation}
In the local Markovian limit, the above equation provides the coupled fluctuation–dissipation relation. A nonzero off-diagonal coefficient therefore requires a corresponding cross correlation between the stochastic noise items. These correlations encode the joint thermal forcing of the two order parameters and must be retained in a numerical calculation of the coupled equations.

The coupled Langevin equations in Eqs.~\eqref{eq:double L1} and~\eqref{eq:double L2} describe the response of two interacting fields to the thermal environment.
In addition to the effect of the effective potential $V_{\rm eff}$, their evolution involves a matrix of dissipative coefficients and correlated stochastic fluctuations. The interaction terms $\eta_{\phi\chi}\dot{\chi}$ and $\eta_{\chi\phi}\dot{\phi}$ couple the dissipative response of the two fields, the motion of either background contributes to the frictional force acting on the other. Eq.~\eqref{eq:dissipation coefficients} further shows that $\eta_{\phi\chi}=\eta_{\chi\phi}\propto \phi\chi$, so these entries vanish when one of the fields is zero. The interaction ingredients can nevertheless affect the diagonal coefficients through the additional fluctuation items and the thermal masses and widths entering the kernels. In the decoupling limit $\lambda_{hs}\to0$, the dissipative coefficients and noise correlations vanish, and the two equations reduce to independent single-field equations. Accounting for the different conventions, the identification $\lambda=6 \lambda_h$ gives
\begin{equation}
    \eta_{\phi\phi} \to 36 \lambda_h^2\phi^2\gamma_{hh}=\eta_1 \phi^2
\end{equation}
with an analogous expression for the $\chi$ sector. However, in order to realize a two-step first-order PTs, we need to choose a nonzero $\lambda_{hs}$.

\section{Summary}
\label{sec:conclusion}

Within the framework of real-time finite-temperature field theory, we employ the CTP formalism to perform a systematic perturbative analysis of nonequilibrium dynamics relevant to first-order electroweak PTs. By computing the real-time effective action, we derive the corresponding stochastic equations of motion for both a single-field $\phi^6$ model and a two-field model. The Langevin equations incorporate both dissipative effects and stochastic fluctuations induced by the thermal bath, with the corresponding dissipation coefficients and noise correlation functions. These results provide a theoretical framework for studying the real-time evolution of scalar fields during first-order PTs. A natural next step is to extend the present analysis by incorporating gauge and fermionic degrees of freedom, which are essential for realistic electroweak theories such as the Standard Model Effective Field Theory and the Standard Model extended by scalar~\cite{Prokopec:2003pj,Prokopec:2004ic,Kainulainen:2021oqs,Ramsey:1997fz}. Gauge and fermionic interactions can generate additional contributions to the effective action, including dissipation and stochastic fluctuations. A systematic treatment of these effects within the CTP formalism will be pursued in future work~\cite{Yamaguchi:1996dp}.

Based on the obtained Langevin equation, we can calculate the bubble nucleation rate. By performing real-time stochastic lattice simulations, one can follow the evolution of thermal field configurations and extract the nucleation rate from the probability of transitions out of the metastable phase~\cite{Berera:2019uyp,Carosi:2024lop}. Previous studies have investigated thermal bubble nucleation using stochastic field dynamics and numerical simulations~\cite{Batini:2023zpi}. Such an approach provides a dynamical description of bubble nucleation that complements the conventional calculation based on the finite-temperature effective potential and Euclidean bounce solutions, also provides an important input for future studies of stochastic gravitational wave and electroweak baryogenesis~\cite{Bergner:2003id,Riotto:1998zb}.

Nevertheless, the Langevin description is derived under the quasiadiabatic and Markovian approximations, together with the use of dressed propagator with finite decay width. The explicit high temperature expressions further employ the narrow width approximation. Reducing the noise kernels to white noise requires their correlation times to be short compared with the background evolution time. While this framework captures dissipative effects and stochastic fluctuations of the scalar background field, it cannot yet self-consistently describe full bubble wall dynamics or nonequilibrium particle transport. A complete description requires simultaneous evolution of the background field and particle correlation functions. The two-particle-irreducible (2PI) effective action and associated Kadanoff–Baym (K-B) equations provide a systematic framework, incorporating nonlocal self-energies and non-equilibrium transport~\cite{Cornwall:1974vz,Berges:2004yj}. Recent studies have applied these methods to investigate nonequilibrium dynamics relevant to cosmological PTs and bubble wall evolution~\cite{Ai:2025bjw,Banik:2026ljf}. Furthermore, the 2PI and K-B formalism is also relevant to nonequilibrium dark matter production, where quantum transport and particle interactions may affect the resulting dark matter abundance~\cite{Kainulainen:2024etd,Becker:2023vwd}. Extending the present analysis along these lines would provide a more complete description of the nonequilibrium dynamics of cosmological first-order PTs.

In summary, we establish a connection between microscopic finite-temperature field theory and the real-time stochastic dynamics of cosmological PTs. The derived Langevin equations provide a theoretical basis for investigating bubble nucleation through real-time simulations. Further extensions incorporating gauge and fermionic contributions, together with more complete nonequilibrium treatments based on the 2PI and K-B formalism, would help clarify the dynamical processes underlying electroweak PTs and their implications for gravitational waves, electroweak baryogenesis, and dark matter phenomenology.

\section*{Acknowledgments}

This work is supported by the National Natural Science Foundation of China (NSFC) under Grants Nos. 12322505, 12547101, and 12675130. L.B. also acknowledges Chongqing Natural Science Foundation under Grant
No. CSTB2024NSCQ-JQX0022 and
Chongqing Talents: Exceptional Young Talents Project No. cstc2024ycjh-bgzxm0020.
	
\appendix
\section{Perturbative Loop Expansion for $\Gamma[\varphi]$}\label{appendix:Gamma}
Here, we outline the detailed derivation of the perturbative loop expansion for the effective action $\Gamma[\varphi]$. Employing the background field method~\cite{Jackiw:1974cv}, we decompose the quantum scalar field into a classical background configuration $\phi$ and a quantum fluctuation $\eta$, such that set $\phi\to\varphi+\eta$. In the presence of an external source $J$, the classical action is expanded as
\begin{equation}\label{eq:Svarphieta}
    S[\varphi+\eta,J]=\int_c \diff^4 x \left\{ \frac{1}{2}(\partial_\mu \varphi)^2+(\partial_\mu \eta)^2-V(\varphi+\eta)+J (\varphi+\eta) \right\}\;,
\end{equation}
where the scalar potential $V(\phi)$ is expanded around the background field as
\begin{align}\label{eq:expandedV}
	V(\varphi+\eta)=V(\varphi)+V'(\varphi)\eta+\frac{1}{2}V''(\varphi)\eta^2+\frac{1}{3!}V^{(3)}(\varphi)\eta^3+\frac{1}{4!}V^{(4)}(\varphi)\eta^4+\frac{1}{5!}c_6\varphi\eta^5+\frac{1}{6!}c_6\eta^6\;,
\end{align}
where 
\begin{align}
	V'(\varphi)&=m^2\varphi+\frac{\lambda}{3!}\varphi^3+\frac{c_6}{5!}\varphi^5\nn\;,\\
	V''(\varphi)&=m^2+\frac{\lambda}{2}\varphi^2+\frac{c_6}{4!}\varphi^4\nn\;,\\
	V^{(3)}(\varphi)&=\lambda \varphi+\frac{c_6}{3!}\varphi^3\nn\;,\\
	V^{(4)}(\varphi)&=\lambda+\frac{c_6}{2}\varphi^2\;.
\end{align}
Next, applying the least action principle, the condition $\left.\delta S/\delta\phi \right|_{\phi=\varphi}=0$ yields the classical field equation for the background configuration $\varphi$:
\begin{equation}
    \partial_\mu \partial^\mu \varphi+V^\prime(\varphi)=J\,.
\end{equation}
Substituting this classical equation into Eq.~\eqref{eq:Svarphieta} cancels all terms linear in the fluctuation field $\eta$. Finally, upon setting the external source $J\to 0$ and integrating out the fluctuation field $\eta$, we obtain the loop expansion for the effective action $\Gamma[\varphi]$:
\begin{align}\label{eq:GammaW1}
	\Gamma[\varphi]=- i \ln \bigg[ \int_c \mathcal{D}\eta \exp{i S[\varphi]+i \int_c\diff^4x \Big( \frac{1}{2}(\partial_\mu \eta)^2-\frac{1}{2}m^2\eta^2+ \mathcal{L}_{\mathrm{int}} \Big)} \bigg]\;,
\end{align}
where the interacting Lagrangian for the fluctuation is given by
\begin{align}
    \mathcal{L}_{\mathrm{int}}=-\frac{1}{2}\lp \frac{\lambda}{2}+\frac{c_6}{4!}\varphi^4\rp\eta^2 -\frac{1}{3!}\lp\lambda \varphi+\frac{c_6}{3!}\varphi^3 \rp \eta^3 -\frac{1}{4!}\lp\lambda+\frac{c_6}{2}\varphi^2 \rp \eta^4-\frac{c_6}{5!}\varphi\eta^5-\frac{c_6}{6!}\eta^6\;.
\end{align}
This expression serves as the starting point for the perturbative loop expansion used in this work.

To proceed, we separate the interaction terms from the free part. The free action $S_0$ of the fluctuation and the interaction $S_{\mathrm{int}}$ are defined as
\begin{equation}
    S_0[\eta]=\int_c\diff^4x \left[\frac{1}{2}(\partial_\mu \eta)^2-\frac{1}{2}m^2\eta^2\right], \quad S_\mathrm{int}=\int_c\diff^4x \mathcal{L}_\mathrm{int}\;.
\end{equation}
Eq.~\eqref{eq:GammaW1} simplifies to
\begin{equation}\label{eq:GammaW2}
    e^{i \Gamma[\varphi]}=e^{i S[\varphi]}\int_c\mathcal{D}\eta e^{i(S_0+S_\mathrm{int})}\;.
\end{equation}
We next evaluate the free Gaussian path integral, denoted as $Z_0$:
\begin{equation}
    Z_0=\int_c \mathcal{D}\eta e^{i S_0[\eta]}=\int_c \mathcal{D}\eta \exp\left\{ i \int_c\diff^4x \frac{1}{2}\eta(-\partial^2-m^2)\eta\right\}\;.
\end{equation}
In momentum space, the inverse propagator satisfies $i\widetilde{D}^{-1}=p^2-m^2$, which allows us to rewrite the Gaussian integral as
\begin{equation}
    Z_0=\int_c \mathcal{D}\eta\exp\left\{i \int_c\diff^4x \frac{1}{2}\eta(i\widetilde{D}^{-1})\eta \right\}\;.
\end{equation}
Evaluating this functional integral yields a well-known result relating the free path integral to the functional determinant of the field operator:
\begin{equation}
    \int_c\mathcal{D}\eta e^{is_0}=\text{const}\times [\det(\partial^2+m^2)]^{-1/2}\;,
\end{equation}
which is identified as a functional determinant.

We then introduce the standard convention for path-integral expectation values,
\begin{equation}
    \braket{F(\eta)}_0 \equiv \frac{\int \mathcal{D}\eta F(\eta)e^{iS_0}}{\int \mathcal{D}\eta e^{iS_0}}\;.
\end{equation}
Substituting this into Eq.~\eqref{eq:GammaW2}, we obtain
\begin{equation}
    e^{i\Gamma[\varphi]}=e^{iS[\varphi]}Z_0\braket{e^{i S_{\mathrm{int}}}}_0\;.
\end{equation}
Taking the logarithm on both sides, the effective action $\Gamma[\varphi]$ decomposes into
\begin{align}\label{eq:Gammaresult}
	\Gamma[\varphi]&=S[\varphi]-i \ln{Z_0}-i\braket{e^{i S_{\mathrm{int}}}}_0\nn\\
	&=S[\varphi]+\frac{i}{2} \Tr \ln{(\partial^2+m^2)}-i \ln \braket{e^{i S_{\mathrm{int}}}}_0\;.
\end{align}
This expression explicitly reveals that the effective action consists of the classical action $S[\varphi]$, the one-loop correction given by the functional determinant, and the connected contributions from all higher-order interactions.
Finally, to extract these higher-order contributions, we employ the cumulant (cluster) expansion for the last term: 
\begin{equation}\label{eq:cluster expansion}
    \ln\braket{e^{X}}_0=\sum_{n=1}^{\infty}\frac{1}{n!}\braket{X^n}_c\;,
\end{equation}
where $X\equiv iS_\mathrm{int}$, and the subscript $c$ denotes the connected parts of the path integrals.
Based on the cluster expansion in Eq.~\eqref{eq:cluster expansion}, the connected contribution from the interactions to the effective action can be explicitly expanded as
\begin{equation}
	-i \ln \braket{e^{i S_{\mathrm{int}}}}_0=-i \sum_{n=1}^{\infty}\frac{i^n}{n!}\braket{S_{\mathrm{int}}^n}_{0,c}\;.
\end{equation}

For the $\phi^6$ model under consideration, the decomposition into the background $\varphi$ and the fluctuation $\eta$ gives rise to three distinct interaction vertices derived from the potential in Eq.~\eqref{eq:expandedV}:
\begin{align}\label{Vertex:phi6}
	-\frac{\lambda}{4}\varphi^2\eta^2&\;,\quad -\frac{c_6}{48}\varphi^4 \eta^2\;,\quad -\frac{\lambda}{6}\varphi \eta^3\;,\quad -\frac{c_6}{36}\varphi^3 \eta^3\nn\;,\\
	-\frac{\lambda}{4!}\eta^4&\;,\quad -\frac{c_6}{48}\varphi^2\eta^4\;,\quad -\frac{c_6}{5!}\varphi\eta^5\;,\quad -\frac{c_6}{6!}\eta^6\;.
\end{align}
These vertices account for all possible interactions ranging from two-point to six-point couplings, characterized by powers of $\varphi$ and $\eta$.

In this work, we truncate the perturbative expansion at the two-loop level. Retaining terms up to $O(\lambda^2)$ and $O(c_6)$, the relevant connected Feynman diagrams contributing to the effective potential are depicted in Eq.~\eqref{eq:action picture}.

\section{The calculation skill in CTP propagators at higher-order loop terms}
\label{appendix:CTP skill}

In this Appendix, we summarize the closed-time-path (CTP) identities useful for evaluating higher-loop contributions to the effective action. These identities allow the contour-index structure to be reduced systematically and make explicit the separation between causal response and fluctuation kernels.

We defines the physical field $\varphi_c$ and the response field $\varphi_\Delta$:
\begin{equation}
	\varphi_+=\varphi_c+\frac{1}{2}\varphi_\Delta\;,\qquad \varphi_-=\varphi_c-\frac{1}{2}\varphi_\Delta\;.
\end{equation}
to substitute these into the effective action components, we need the following expansions:
\begin{equation}
	\begin{aligned}
		\varphi_+^2 - \varphi_-^2 &= 2\varphi_c\varphi_\Delta\;, \\
		\varphi_+^3 - \varphi_-^3 &= 3\varphi_c^2\varphi_\Delta + \frac{1}{4}\varphi_\Delta^3\;,\\
		\varphi_+^4 - \varphi_-^4 &= 4\varphi_c^3\varphi_\Delta + \varphi_c\varphi_\Delta^3\;,\\
		\varphi_\pm^2 = \varphi_c^2 &\pm \varphi_c\varphi_\Delta + \frac{1}{4}\varphi_\Delta^2\;.
	\end{aligned}
\end{equation}
To simplify these, let $X(x)=\varphi_c^2(x)+\frac{1}{4}\varphi_\Delta^2(x)$ and $Y(x)=\varphi_c(x)\varphi_\Delta(x)$, such that $\varphi_\pm^2=X\pm Y$.

Let $\tau=t-t^\prime$, and consider a generic product of CTP propagators carrying the same contour indices,
\begin{equation}
    \Pi^{ab}(\tau)\equiv \prod_{j=1}^{N}G^{ab}_j(\tau), \qquad a,b=\pm\;, 
\end{equation}
where $j$ collectively labels the momentum, mass, and field species of each internal line.
The four CTP propagators can be written in the Wightman functions as 
\begin{equation}
    \begin{aligned}
        G^{++}(\tau) &= G^>(\tau)\theta(\tau) + G^<(\tau)\theta(-\tau)\;,\\
		G^{--}(\tau) &= G^>(\tau)\theta(-\tau) + G^<(\tau)\theta(\tau)\;,
    \end{aligned}
\end{equation}
and
\begin{equation}
G^{+-}(\tau) = G^>(\tau),\quad G^{-+}(\tau) = G^<(\tau)\;.
\end{equation}
Using 
\begin{equation}
    \theta(\tau)\theta(-\tau)=0\;, \quad [\theta(\pm\tau)]^{N}=\theta(\pm \tau)\;,
\end{equation}
one immediately finds that an generalized product of these propagetors inherits the same contour structure as an individual propagetor,
\begin{equation}
    \Pi^{++}=\theta(\tau)\Pi^{>}+\theta(-\tau)\Pi^{<}\;, \quad \Pi^{--}=\theta(-\tau)\Pi^{>}+\theta(\tau)\Pi^{<}\;,
\end{equation}
together with
\begin{equation}
    \Pi^{+-}=\Pi^{>}\;, \qquad \Pi^{-+}=\Pi^{<}\;,
\end{equation}
where
\begin{equation}
    \Pi^{>}=\prod_{j=1}^{N}G_J^{>}\;.\qquad \Pi^{<}=\prod_{j=1}^{N}G_J^{<}\;.
\end{equation}
When the time dependence is implicit, we write $\Pi^<(\tau)$ and $\Pi^>(\tau)$ as $\Pi^<$, $\Pi^>$ for brevity.
We take into account that the propagator satisfies $\Pi^{<}(\tau)=[\Pi^{>}(-\tau)]^{*}$. Therefore, the following four particularly useful CTP combinations can be constructed:
\begin{align}
    &\Pi^{++} - \Pi^{+-} - \Pi^{-+} + \Pi^{--} = 0\;,\label{eq:Kernel1}\\
    &\Pi^{++} + \Pi^{+-} - \Pi^{-+} - \Pi^{--} = 2\theta(\tau)(\Pi^> - \Pi^<) = 4i \theta(\tau) \Im(\Pi^{++})\;,\label{eq:Kernel2}\\
    &\Pi^{++} - \Pi^{+-} + \Pi^{-+} - \Pi^{--} = -2\theta(-\tau)(\Pi^> - \Pi^<) = 4i \theta(-\tau) \Im(\Pi^{++})\;,\label{eq:Kernel3}\\
    &\Pi^{++} + \Pi^{+-} + \Pi^{-+} + \Pi^{--} = 2(\Pi^> + \Pi^<) = 4 Re(\Pi^{++})\label{eq:Kernel4}\;.
\end{align}
Equation~\eqref{eq:Kernel1} ensures the cancellation of contributions containing only classical fields, consistently with the normalization property of the CTP effective action. Equations~\eqref{eq:Kernel2} and~\eqref{eq:Kernel3} isolate the retarded and advanced components respectively, that expressions are exactly the retarded and advanced component:$\Pi^R=\theta(t)(\Pi^> -\Pi^<)$ and $\Pi^A=\theta(-\tau)(\Pi^< -\Pi^>)$, and its imaginary part produces dissipation. Odd in $\varphi_\Delta$ is the disspation item, even in $\varphi_\Delta$ is the noise item, whereas Equation~\eqref{eq:Kernel4} selects the real correlation part.
Note that the above identities are independent of the number of propagators in $\Pi^{ab}$ hence the same CTP reduction method applies regardless of whether two, three, or more internal propagators are present.

Now, consider a generic two-local contribution of the form
\begin{align}\label{eq:CTP generic form}
    \mathcal{I}(x,x^{\prime})=F_+ \Pi^{++}(x,x^\prime) F_+ - F_+ \Pi^{+-}(x,x^\prime) F_- -F_- \Pi^{-+}(x,x^\prime) F_+ +F_- \Pi^{--}(x,x^\prime) F_-
\end{align}
For quadratic insertions,$F_{\pm}=X(x)\pm Y(x)$. Then Eq.~\eqref{eq:CTP generic form} can be rearranged as 
\begin{align}
    \mathcal{I}(x,x^{\prime})&=X(x)X(x^\prime) \lp \Pi^{++}-\Pi^{+-}-\Pi^{-+}+\Pi^{--} \rp + X(x)Y(x^\prime) \lp \Pi^{++}+\Pi^{+-}-\Pi^{-+}-\Pi^{--} \rp \nn\\
    &+Y(x)X(x^\prime) \lp \Pi^{++}-\Pi^{+-}+\Pi^{-+}-\Pi^{--} \rp+Y(x)Y(x^\prime)\lp \Pi^{++}+\Pi^{+-}+\Pi^{-+}+\Pi^{--} \rp\;.
\end{align}
Substituting Eqs.~\eqref{eq:Kernel1}-\eqref{eq:Kernel4}, we obtain
\begin{align}
    \mathcal{I}(x,x^{\prime})=&4 i X(x) Y(x^\prime)\theta(t-t')\Im \Pi^{++}(x,x')+4 i Y(x) X(x^\prime)\theta(t'-t)\Im \Pi^{++}(x,x')\nn\\
    &+4Y(x)Y(x')\Re \Pi^{++}(x,x')\;.
\end{align}
In order to manifest the law of causality, we perform a change of integration variables, for the second term, perform the exchange $x \leftrightarrow x'$, and simultaneously redefine the internal momentum so the it remains invariant. In this way, the second term can be transformed into the first term. Therefore, the first two terms can be combined into a single explicit expression.
\begin{align}
    \int\diff^4x\int\diff^4x' \mathcal{I}(x,x^{\prime})=&8 i\int\diff^4x\int\diff^4x' X(x)Y(x')\theta(t-t')\Im \Pi^{++}(x,x')\nn\\
    &+4 \int\diff^4x\int\diff^4x' Y(x)Y(x')\Re \Pi^{++}(x,x')\;.
\end{align}
This method allows us to fully compute the higher-order contributions in CTP theory.

\section{Detailed loop expansion of the effective action for the two-field model}\label{appendix:two-field model}
In this appendix, we provide the detailed derivation of the effective action $\Gamma[\phi,\chi]$ for the two-field model introduced in Sec.~\ref{sec:two field model}. Since the cumulant expansion and the relevant CTP propagator identities have already been established in the single-field analysis, we do not repeat their general derivation. Instead, we focus on the different internal-particle channels generated by the two fluctuation fields. Decomposing the quantum fields into classical backgrounds and quantum fluctuations, $H=\phi+h$ and $S=\chi+s$, the action $S[H,S]$ can be expanded as a Taylor series around the classical fields $\phi$ and $\chi$:
\begin{align}
S[\phi+h,\chi+s]=&S[\phi,\chi]+\int \diff^4 x\Big(\frac{\delta S}{\delta \phi(x)}h(x)+\frac{\delta S}{\delta \chi(x)}s(x)\Big)\nn\\
&+\frac{1}{2}\int\diff^4x\diff^4y\begin{pmatrix}h(x) & s(x)\end{pmatrix}\frac{\delta^2 S}{\delta \Phi^2}(x,y)\begin{pmatrix}h(y) \\ s(y)\end{pmatrix}+S_{\mathrm{int}}\;,
\end{align}
where $\Phi$ collectively denotes $\phi$ and $\chi$, and $S_\mathrm{int}$ contains the interaction terms of cubic and higher orders in the fluctuations.

For the perturbative prescription adopted in this work, it is useful to separate the background-independent quadratic terms form the remaining background-dependent interactions. The expansion of the potential can be written as
\begin{align}
	V(\phi+h,\chi+s)=&V(\phi,\chi)+\frac{\delta V}{\delta \phi}h+\frac{\delta V}{\delta \chi}s+\frac{1}{2}\frac{\delta^2 V}{\delta \phi^2}h^2+\frac{1}{2}\frac{\delta^2 V}{\delta \chi^2}s^2+\frac{\delta^2 V}{\delta\phi \delta \chi}h s+V_{3}+V_{4}\nn\\
	=&V(\phi,\chi)+\Big(m_h^2 \phi+\lambda_h \phi^3+\frac{1}{2}\lambda_{hs}\phi\chi^2\Big)h+\Big(m_s^2 \phi+\lambda_s \chi^3+\frac{1}{2}\lambda_{hs}\phi^2\chi\Big)s\nn\\
	&+\frac{1}{2}\Big(m_h^2+3\lambda_h \phi^2+\frac{1}{2}\lambda_{hs}\chi^2 \Big)h^2+\frac{1}{2}\Big( m_s^2+3\lambda_s \chi^2+\frac{1}{2}\lambda_{hs}\phi^2 \Big)s^2\nn\\
    &+\lambda_{hs}\phi\chi hs+V_{3}+V_4\;,
\end{align}
where
\begin{align}
    V_{3}&=\lambda_h \phi h^3+\lambda_s \chi s^3+\frac{\lambda_{hs}}{2}\phi h s^2+\frac{\lambda_{hs}}{2}\chi h^2 s\;,\nn\\
    V_{4}&=\frac{\lambda_h}{4}h^4+\frac{\lambda_s}{4}s^4+\frac{\lambda_{hs}}{4}h^2s^2\;.
\end{align}
The terms linear in the fluctuations are canceled by the source terms entering the Legendre transformation.We therefore decompose the fluctuation action on the closed time contour as
\begin{equation}
    S_{\mathrm{fluct,c}}=S_{0,c}+S_{\mathrm{2,c}}++S_{\mathrm{3,c}}++S_{\mathrm{4,c}}\;,
\end{equation}
where the background-independent Gaussian action is chosen as
\begin{equation}
    S_{0,c}=\int_c \diff^4 x \Big[ \frac{1}{2}(\partial_\mu h)^2-\frac{1}{2}m_h^2 h^2+\frac{1}{2}(\partial_\mu s)^2-\frac{1}{2}m_s^2 s^2\Big]\;.
\end{equation}
All bacground-dependent quadratic terms are treated as additionanal interaction vertices,
\begin{equation}
    S_{\mathrm{2,c}}=-\int_c\diff^4 x \Big[ \frac{3}{2}\lambda_h\phi^2 h^2+\frac{1}{4} \lambda_{hs}\chi^2 h^2+\frac{3}{2}\lambda_s\chi^2s^2+\frac{1}{4}\lambda_{hs}\phi^2s^2+\lambda_{hs}\phi\chi hs\Big]\;,
\end{equation}
\begin{equation}
    S_{\mathrm{3,c}}=-\int_c\diff^4 x \Big[ \lambda_h \phi h^3+\frac{1}{2}\lambda_{hs} \phi h s^2+\lambda_s\chi s^3+\frac{1}{2}\lambda_{hs}\chi h^2 s\Big]\;,
\end{equation}
\begin{equation}
    S_{\mathrm{4,c}}=-\int_c\diff^4 x \Big[ \frac{1}{4}\lambda_h h^4+\frac{1}{4}\lambda_s s^4+\frac{1}{4}\lambda_{hs}h^2 s^2\Big]\;.
\end{equation}
The purpose of this work is to evaluate the background-dependent part of the closed-time-path effective action. We perform a simultaneous expansion in the scalar couplings and in the number of loops, retaining terms through $\mathcal{O}(\lambda^2)$ and up to two-loop order. Throughout the calculation, we assume
\begin{equation}
    \lambda_h \sim \lambda_s \sim \lambda_{hs} \sim \mathcal{O}(\lambda)\;.
\end{equation}
At this order, the relevant contributions can be depicted in Figure~\ref{fig:double-field-action} .

Next, we provide the explicit diagrammatic evaluation of the two- field effective action introduced in Sec.~\ref{sec:two field model}.
We first define
\begin{equation}\label{eq:F1}
    F_1[Q,\Psi]:=\int \diff^4 x\int \frac{\diff^3 q}{(2 \pi)^3}\bigg\{ Q_+(x) G_{\Psi}^{++}(\bm{q},0)-Q_-(x) G_{\Psi}^{--}(\bm{q},0)\bigg\}\;,
\end{equation}
where $\phi$ denote the background field and $\Psi=h,s$. Using Eq.~\eqref{eq:F1}, the one row contribution in Figure~\ref{fig:double-field-action} is
\begin{equation}
    \Gamma_2=C_{\phi\phi hh}F_1[\phi^2,h]+C_{\phi\phi ss}F_1[\phi^2,s]+C_{\chi\chi hh}F_1[\chi^2,h]+C_{\chi\chi ss}F_1[\chi^2,s]\;.
\end{equation}

For the second row in Figure~\ref{fig:double-field-action}, let’s define its characteristic integral
\begin{align}
F_2[Q_1,Q_2,\Psi_1,\Psi_2]:=&\int d^4 x d^4 x^\prime \int \frac{d^3 k}{(2 \pi)^3} e^{i \bm{k}\cdot(\bm{x}-\bm{x^\prime})} \int \frac{d^3 q}{(2 \pi)^3} \bigg[ Q_{1+} (x) G_{\Psi_1}^{++}(\bm{q},t-t') \nn\\
&G_{\Psi_2}^{++}(\bm{q}-\bm{k} , t-t') Q_{2+} (x')- Q_{1+} (x) G_{\Psi_1}^{+-} (\bm{q} , t-t') G_{\Psi_2}^{+-}(\bm{q} -\bm{k} ,t-t')\nn\\
&\times Q_{2-} (x')- Q_{1-} (x) G_{\Psi_1}^{-+}(\bm{q} , t-t')G_{\Psi_2}^{-+}(\bm{q} -\bm{k} , t-t')Q_{2+} (x') \nn\\
&+ Q_{1-} (x) G_{\Psi_1}^{--}(\bm{q}, t-t')G_{\Psi_2}^{--}(\bm{q} -\bm{k} ,t-t') Q_{2-} (x')\bigg]\;,
\end{align}
we have 
\begin{align}
	\Gamma_{22}=&iC_{\phi\phi h h}^2F_2[\phi^2,\phi^2,h,h]+iC_{\phi\phi s s}^2F_2[\phi^2,\phi^2,s,s]\nn\\
	&+i2 C_{\phi\phi hh} C_{\chi\chi hh}F_2[\phi^2,\chi^2,h,h]+i2C_{\phi\phi ss} C_{\chi\chi ss}F_2[\phi^2,\chi^2,s,s]\nn\\
	&+iC_{\chi\chi h h}^2F_2[\chi^2,\chi^2,h,h]+iC_{\chi\chi s s}^2F_2[\chi^2,\chi^2,s,s]\nn\\
    &+\frac{i}{2}C_{\phi\chi hs}^2F_2[\phi\chi,\phi\chi,h,s]\;.
\end{align}
Next, let’s define the third type of characteristic integral,
\begin{align}
	F_3[Q,\Psi_1,\Psi_2,\Psi_3]:=&\int \frac{\diff^3 k}{(2 \pi)^3}\frac{1+2 n(\omega)}{2 \omega(\bm{k})}\int \diff^4 x\int \diff t' \frac{\diff^3 q}{(2 \pi)^3}\bigg\{Q_+(x)G_{\Psi_1}^{++}(\bm{q},t-t')G_{\Psi_2}^{++}(\bm{q},t-t')\nn\\
	&-Q_+(x)G_{\Psi_1}^{+-}(\bm{q},t-t')G_{\Psi_2}^{+-}(\bm{q},t-t')-Q_-(x)G_{\Psi_1}^{-+}(\bm{q},t-t')G_{\Psi_2}^{-+}(\bm{q},t-t')\nn\\
	&+Q_-(x)G_{\Psi_1}^{--}(\bm{q},t-t')G_{\Psi_2}^{--}(\bm{q},t-t')\bigg\}\;,
\end{align}
where
\begin{equation}
	\int\frac{\diff^3 k}{(2 \pi)^3}\frac{1+2 n(\omega)}{2 \omega(\bm{k})}=\int\frac{\diff^3 k}{(2 \pi)^3}G_{\Psi_3}^{++}(\bm{k},0)=\int\frac{\diff^3 k}{(2 \pi)^3}G_{\Psi_3}^{--}(\bm{k},0)\;.
\end{equation}
We have
\begin{align}
\Gamma_{24}=&i12 C_{\phi\phi hh}C_{hhhh}F_3[\phi^2,h,h,h]+i2C_{\phi\phi hh}C_{hhss}F_3[\phi^2,h,h,s]+i12C_{\phi\phi ss}C_{ssss}F_3[\phi^2,s,s,s]\nn\\
&+i2C_{\phi\phi ss}C_{sshh}F_3[\phi^2,s,s,h]+i12 C_{\chi\chi hh}C_{hhhh}F_3[\chi^2,h,h,h]+i2C_{\chi\chi hh}C_{hhss}F_3[\chi^2,h,h,s]\nn\\
&+i2C_{\chi\chi ss}C_{sshh}F_3[\chi^2,s,s,h]+i12C_{\chi\chi ss}C_{ssss}F_3[\chi^2,s,s,s]
\end{align}
We last define 
\begin{align}
	F_4[Q,\Psi_1,\Psi_2,\Psi_3]:=&\int \diff^4 x \diff^4 x' \int \frac{\diff^3 k}{(2 \pi)^3} e^{i \bm{k}.(\bm{x}-\bm{x}')}\int \frac{\diff^3 q_1}{(2 \pi)^3} \frac{\diff^3 q_2}{(2 \pi)^3} \frac{\diff^3 q_3}{(2 \pi)^3} \delta(\bm{k}-\bm{q}_1-\bm{q}_2-\bm{q}_3)\nn\\
	&\times \bigg\{Q_+(x)G_{\Psi_1}^{++}(\bm{q}_1,t-t')G_{\Psi_2}^{++}(\bm{q}_2,t-t')G_{\Psi_3}^{++}(\bm{q}_3,t-t')Q_+(x')\nn\\
	&-Q_+(x)G_{\Psi_1}^{+-}(\bm{q}_1,t-t')G_{\Psi_2}^{+-}(\bm{q}_2,t-t')G_{\Psi_3}^{+-}(\bm{q}_3,t-t')Q_-(x')\nn\\
	&-Q_-(x)G_{\Psi_1}^{-+}(\bm{q}_1,t-t')G_{\Psi_2}^{-+}(\bm{q}_2,t-t')G_{\Psi_3}^{-+}(\bm{q}_3,t-t')Q_+(x')\nn\\
	&+Q_-(x)G_{\Psi_1}^{--}(\bm{q}_1,t-t')G_{\Psi_2}^{--}(\bm{q}_2,t-t')G_{\Psi_3}^{--}(\bm{q}_3,t-t')Q_-(x')\bigg\}\;,
\end{align}
and we have
\begin{align}
\Gamma_{33}=i3C_{\phi hhh}^2F_4[\phi,h,h,h]+iC_{\phi hss}^2F_4[\phi,h,s,s]+iC_{\chi hhs}^2F_4[\chi,h,h,s]+i3C_{\chi sss}^2F_4[\chi,s,s,s]\;.
\end{align}

In the above, $C_{\Psi_i \Psi_j}$ denote vertex coefficients for generic filed $\Psi_i$, and an exhaustive list of required vertex coefficients read
\begin{align}
C_{\phi\phi hh}&=-\frac{3}{2}\lambda_h\,,& &C_{\chi\chi ss}=-\frac{3}{2}\lambda_s\;,& &C_{\phi\phi ss}=C_{\chi\chi hh}=-\frac{1}{4}\lambda_{hs}\,,\\
C_{\phi hhh}&=-\lambda_h\,,& &C_{\chi sss}=-\lambda_s\;,& &C_{\phi hss}=C_{\chi hhs}=-\frac{1}{2}\lambda_{hs}\,,\\
C_{hhhh}&=-\frac{1}{4}\lambda_h\,,& &C_{ssss}=-\frac{1}{4}\lambda_s\;,& &C_{hhss}=-\frac{1}{4}\lambda_{hs}\,.
\end{align}

Collecting the four classes of diagrams, the loop contribution to the two-field CTP effective action is
\begin{equation}
    \Gamma_{\rm loop}=\Gamma_2+\Gamma_{22}+\Gamma_{24}+\Gamma_{33}+\mathcal{O}(\lambda^3)\;.
\end{equation}

\section{the main integral}

In this appendix, We concentrate in the master integrals involved in our calculation. We begin with the two types of integrals in the single-field case, whose treatment can be generalized to the two-field model.
First, the function $I(m)$, the basic 1-loop structure appearing within the 2-loop result, is denoted by
\begin{equation}
    I(m)=\int\frac{\diff^3 k}{(2\pi)^3}\frac{1+2n(\omega)}{2 \omega(\bm{k})}\;,
\end{equation}
where $n(\omega)=\lp e^{\beta \omega}-1 \rp^{-1}$ is the Bose distribution and $\omega \equiv \omega(\bm{k})=\sqrt{\bm{k}^2+m^2}$ is the free particle energy.
In the high-T limit, we have
\begin{align}
    I(m)=\frac{T^2}{12}-\frac{m T}{4 \pi}-\frac{2 m^2}{(4\pi)^2}\ln\lp \frac{\bar{\mu}e^{\gammaE}}{4\pi T}\rp+\mathcal{O}(\frac{m^2}{T^2})\;.
\end{align}

For the Second function $K(m)$, we define 
\begin{equation}\label{eq:function K}
    K(m)=\int_{-\infty}^{t}\diff t'\int\frac{\diff^3 q}{(2\pi)^3}\Im\Big[G_{\phi}^{++}(\bm{q},t-t')\Big]^2\;.
\end{equation}
Let $\tau=t-t'$, the physical propagator $G_{\phi}^{++}(\bm{q},t-t')$ is defined by~\cite{Gleiser:1993ea} 
\begin{equation}
    G_{\phi}^{++}(\bm{q},\tau)=\frac{e^{-\Gamma \tau}}{2\omega(\bm{q})}\bigg[(1+2n)\cos(\omega \tau)+2\beta \Gamma n(1+n)\sin(\omega \tau)-i \sin(\omega\tau) \bigg]\;.
\end{equation}
To solve for $\Im[G^{++}]^2$ in Eq.~\eqref{eq:function K}, we can write $G^{++}$ as
\begin{equation}
    G^{++}=A-i B\;,
\end{equation}
where
\begin{align}
    A&=\frac{e^{-\Gamma \tau}}{2\omega(\bm{q})}\bigg[(1+2n)\cos(\omega \tau)+2\beta \Gamma n(1+n)\sin(\omega \tau)\bigg]\;,\nn\\
    B&=\frac{e^{-\Gamma \tau}}{2\omega(\bm{q})}\sin(\omega\tau)\;.
\end{align}
We can get
\begin{align}
\Im[G^{++}]^2&=\Im[A-iB]^2=-2AB\nn\;,\\
&=-\frac{e^{-2 \Gamma\tau}}{2\omega^2}\sin(\omega\tau)\Big[ (1+2n)\cos(\omega \tau)+2\beta \Gamma n(1+n)\sin(\omega \tau)\Big]\nn\\
&=-\frac{e^{-2 \Gamma\tau}}{2\omega^2}\bigg[ \frac{1+2n}{2}\sin(2\omega\tau)+\beta \Gamma n(1+n)\Big[1-\cos(2 \omega \tau)\Big]\bigg]\;.
\end{align}
For the time integration, we have the following special integral:
\begin{align}
    \int_{0}^{+\infty}\diff\tau &e^{- \Gamma \tau} \sin \omega \tau =\frac{\omega}{\omega^2+\Gamma^2}\nn\;,\\
    \int_{0}^{+\infty}\diff\tau &e^{- \Gamma \tau} \cos \omega \tau =\frac{\Gamma}{\omega^2+\Gamma^2}\nn\;,\\
    \int_{0}^{+\infty}\diff\tau &e^{- \Gamma \tau} =\frac{1}{\Gamma}\;.
\end{align}
We have
\begin{align}
    \int_{0}^{+\infty}\diff\tau \Im[G^{++}]^2=-\frac{1}{2\omega^2}\bigg[\frac{1+2n}{2} \frac{\omega}{2(\omega^2+\Gamma^2)}+\beta \Gamma n(1+n)\Big(\frac{1}{2\Gamma}-\frac{\Gamma}{2(\omega^2+\Gamma^2)}\Big)\bigg]\;.
\end{align}
In the weak coupling limit where $\Gamma \ll m$, we proceed with the expansion, we obtain
\begin{equation}
    \int\frac{\diff^3 q}{(2 \pi)^3}\int_{0}^{+\infty}\diff\tau \Im[G^{++}]^2=\int\frac{\diff^3q}{(2\pi)}\bigg[ -\frac{1}{2\omega^2}\lp\frac{1+2n}{4\omega}+\frac{\beta n(1+n)}{2}\rp\bigg]\;,
\end{equation}
then we can get
\begin{equation}
    K(m)=-\frac{T}{16\pi m}-\frac{1}{32\pi^2}\ln\frac{\bar{\mu}^2}{T^2}+\frac{\ln 4 \pi-\gammaE}{16 \pi^2}\;.
\end{equation}
Next, we summarize the thermal integrals entering the characteristic functions $F_1$-$F_4$ defined in Appendix~\ref{appendix:two-field model}. After separating the background field dependence, the relevant propagator structures can be reduced to a set of thermal kernels. In the following, $I,K$ and $L$ denote the thermal kernels associated with $F_1, F_2$ and $F_4$, respectively, while the contribution to $F_3$ factorizes into the kernels appearing in $F_1$ and $F_2$.
For $F_2$, we define the real-time two propagators kernel
\begin{equation}
   F_2\propto K(m,M)=\int_{-\infty}^{t}dt'\int\frac{\diff^3 q}{(2\pi)^3}\Im\Big[G^{++}_{\Psi_1}(\bm{q},t-t')G^{++}_{\Psi_2}(\bm{q},t-t')  \Big]\;,
\end{equation}
where m and M denote the masses of the fields $\Psi_1$ and $\Psi_2$ respectively. Consequently,
\begin{equation}
    K(m,M)=-\frac{T}{8\pi(m+M)}-\frac{L_b}{32\pi^2}+\frac{(m^2+M^2)\zeta(3)}{256\pi^4T^2}+\cdots\;,
\end{equation}
where $L_b=2\ln\lp\frac{\bar{\mu}}{T}\rp-2\left[\ln 4\pi-\gammaE \right]$.
The thermal structure of $F_3$ does not introduce an additional master integral. Owing to its factorized topology, it can be written as
\begin{equation}
    F_3\propto\mathcal{I}_3(m_1,m_2,m_3)=I(m_3)K(m_1,m_2)\;.
\end{equation}
The characteristic function $F_4$ contains three propagators connecting the two interaction vertices. In the local and zero external momentum limit, we define
\begin{equation}
    F_4\propto L(m_1,m_2,m_3)\equiv\int_0^{\infty}\diff \tau \Im \Big[\prod_{j=1}^{3} \int \frac{\diff^3 q_j}{(2 \pi)^3}G^{++}_{\Psi_j}(\bm{q}_j,\tau) \Big] \delta(\bm{q}_1-\bm{q}_2-\bm{q}_3)\;,
\end{equation}
we have 
\begin{equation}
    L(m_1,m_2,m_3)=-\frac{T^2}{2(4\pi)^2}\Big[ \ln{\frac{\bar{\mu}}{m_1+m_2+m_3}}+\frac{1}{2}\Big]+\cdots\;.
\end{equation}

\bibliographystyle{JHEP}
\bibliography{reference.bib}
\end{document}